\documentclass[twocolumn]{aastex702} 
\usepackage{amsmath}
\usepackage{xcolor}

\shorttitle{Polarization Angle Calibration via Stellar Diffraction}
\shortauthors{Bass \& Johnson}

\begin{document}

\title{A Single-Shot Method for the Absolute Polarization Angle Calibration of CMB Telescopes with Artificial Sky Sources via Stellar Diffraction Encoding}

\author[orcid=0009-0001-1448-315X,gname=Dillon,sname=Bass]{Dillon J. Bass}
\affiliation{University of Virginia, Department of Astronomy, Charlottesville, VA 22904, USA}
\email[show]{ntk5ga@virginia.edu}

\author[orcid=0000-0002-6898-8938,gname=Bradley,sname=Johnson]{Bradley R. Johnson}
\affiliation{University of Virginia, Department of Astronomy, Charlottesville, VA 22904, USA}
\email{bradley.johnson@virginia.edu}

\correspondingauthor{Dillon J. Bass}

\begin{abstract}

Absolute polarization angle calibration of cosmic microwave background (CMB) telescopes at the \(\lesssim0.1^\circ\) level is needed to prevent instrumental mixing of \(E\) modes into \(B\) modes from biasing constraints on primordial gravitational waves and cosmic birefringence. Artificial polarized sky sources like drones, balloons, or satellites can provide calibration references, but transferring their emitted polarization angle from the payload frame to the celestial frame remains a challenge. We propose a calibrator architecture, applicable to any such source, in which a single dual-use grid-optic sets the polarization axis of the emitted millimeter-wave signal and also diffracts starlight into a co-aligned star camera. The resulting diffraction spectra record the grid's orientation relative to the stars, so that it can be measured directly on the sky without fine alignment between the camera and grid-optic. We describe the geometry and physics of this concept and show that it significantly simplifies obtaining the calibration angle the observatory needs, because precise payload attitude is no longer needed and the received polarization is insensitive to small line-of-sight errors. In simulations of 1000 real star fields that include camera--grid misalignment, lens distortion, lateral chromatic aberration, atmospheric refraction, and stellar colors, a single exposure recovers the polarizing grid-optic's orientation in the celestial frame with an uncertainty of \(\simeq0.002^\circ\), fifty times below the \(0.1^\circ\) goal. We also discuss the hardware tolerances a physical calibrator must meet, which future work will test. Our results establish the feasibility of this approach and motivate development of a dual-use grid-optic.

\end{abstract}





\section{Introduction}
\label{sec:intro}

Measurements of the cosmic microwave background (CMB) have become one of the primary tools for studying the contents, geometry, and evolution of the universe. In addition to temperature anisotropies, the CMB contains a linearly polarized component that can be decomposed into even-parity $E$ modes and odd-parity $B$ modes \citep{Zaldarriaga1997, Kamionkowski1997}. While scalar density perturbations generate the dominant temperature and $E$-mode anisotropies, primordial tensor perturbations from inflation are expected to produce a faint curl-like $B$-mode component \citep{Zaldarriaga1997, Kamionkowski1997}. A detection of this primordial signal would directly probe early-universe physics and strongly constrain inflationary models. Current and next-generation CMB experiments, like the Simons Observatory, are therefore designed to place increasingly stringent constraints on the tensor-to-scalar ratio $r$ \citep{Ade2019,Hazumi2020, abazajian_2022}.

However, this measurement is challenging because the expected primordial $B$-mode signal is extremely faint compared to the dominant $E$-mode polarization and other sources of polarized emission. Observed $B$ modes can be produced not only by primordial gravitational waves, but also by gravitational lensing of $E$ modes, polarized Galactic foregrounds, and instrumental systematics \citep{Kamionkowski2016}. One particularly important systematic is uncertainty in the absolute polarization angle of a telescope's detectors---the orientation of their polarization sensitivity relative to celestial coordinates. A global miscalibration $\alpha$ of this angle rotates the measured polarization field, transferring a fraction $\sin^2 2\alpha$ of the $E$-mode power into $B$ modes and producing a spurious $B$-mode signal \citep{ODea2007}. Because the $E$-mode signal is so much larger, even a small angle error can bias constraints on $r$. Instrumental rotations are also observationally degenerate with cosmic birefringence, making independent absolute polarization angle calibration necessary for separating instrumental miscalibration from possible parity-violating physics \citep{Carroll1990, Keating2013, Minami2019}, a question made more pressing by tentative measurements of a birefringence angle $\beta \approx 0.3^\circ$ in Planck data \citep{Minami2020, DiegoPalazuelos2022}. For these reasons, absolute polarization angle calibration has become a central requirement for modern CMB experiments \citep{Bock2006}. Current data limit $r_{0.05} < 0.036$ at 95\% confidence \citep{BK18}, and next-generation experiments aim to reach $\sigma(r) \sim 10^{-3}$, more than an order of magnitude beyond this bound \citep{Ade2019, abazajian_2022}. At that sensitivity, an angle error of a few tenths of a degree is enough to bias $r$ significantly, so the polarization angle must be calibrated to $\mathcal{O}(0.1^\circ)$ or better \citep{Bryan2018, abitbol_2021, vielva_2022}.

Several strategies have been developed to address polarization angle calibration. Self-calibration methods use the expectation that the parity-odd $TB$ and $EB$ spectra vanish in the absence of instrumental rotation, but this approach can absorb or remove real physical signals like cosmic birefringence \citep{Keating2013, Krachmalnicoff2022}. Astrophysical sources provide an external reference, with Tau A commonly used as one of the best polarized calibrators, but their intrinsic polarization properties are known only to finite precision, which is not sufficient for the most stringent upcoming calibration goals \citep{PlanckCollaboration2015, Aumont2020, Johnson2015}. Planets are widely used for beam characterization, but they are effectively unpolarized and therefore hard to use for absolute polarization angle calibration \citep{Weiland2011, PlanckCollaboration2017, Lungu2022}. Dedicated ground-based calibrators, such as the sparse wire grid of the Simons Observatory \citep{Murata2023, Nakata2026} and the rotating polarized source of BICEP3 \citep{Cornelison2022, BK18angles2024}, now approach the required level for small-aperture telescopes, but they reference the polarizer to gravity or to a surveyed surface rather than to the sky, and cannot easily be placed in the far field of large telescopes \citep{Savorgnano2026}.

These limitations have motivated the development of artificial polarized calibration sources that are designed to be bright, highly polarized, and placed in the far field of the instrument under calibration. Proposed and demonstrated approaches include drone-borne sources such as PROTOCALC \citep{Coppi2025} and HoverCal/POLOCALC \citep{Dunner2024, Nati_2017}, as well as space-based concepts such as CalSat \citep{Johnson2015}, LEO-CalSat \citep{Casas2026}, and COSMOCal \citep{Ritacco2024}. These systems aim to replace natural sky sources like Tau A with a precisely engineered polarized emitter, but they introduce a new problem: the polarization angle of the emitted calibration signal is naturally defined in the payload frame, while the CMB experiment needs it in the celestial frame. Converting between these frames generally requires precise laboratory metrology, stable mechanical alignment, source orientation knowledge, payload attitude reconstruction, and a chain of coordinate transformations. At the $\sim 0.1^\circ$ level, these transformations become a central limitation on calibration precision.

For example, PROTOCALC, a W-band drone-based polarized calibrator designed for the Simons Observatory, demonstrates both the promise of artificial polarized sources and the difficulty of transferring the source polarization angle into the celestial frame \citep{Coppi2025}. In this system, the source attitude is reconstructed in flight using GNSS/GPS, gyroscopes, accelerometers, and ground-target photogrammetry, while the relative orientation between the photogrammetry camera and the polarizing wire grid is measured on the ground using a laser-generated diffraction pattern from the grid. The payload's internal camera--grid alignment can be measured at the sub-$0.01^\circ$ level, and the in-flight roll angle can be reconstructed with a reported statistical accuracy of $0.045^\circ$ \citep{Coppi2025}.

A similar diffraction measurement is made on the payload itself by COSMOCal, a calibrator planned for geostationary orbit whose ground prototype has been tested with the NIKA2 camera on the IRAM 30\,m telescope \citep{Ritacco2024, Micheli2026}. There, a camera inside the calibrator images both the laser diffraction pattern from the grid and geo-referenced ground landmarks, which give the camera's roll through photogrammetry \citep{Ritacco2024}. In the IRAM 30\,m test, the calibrator's internal precision reached $\sim 0.03^\circ$, limited mainly by this roll, whose solution depended on which landmarks were used and on how well their positions were known, although near-field and telescope-side systematics kept the comparison with NIKA2 at the few-degree level \citep{Savorgnano2026}. For the space instrument, photogrammetry may no longer be needed if the spacecraft pointing and instrument geometry are independently controlled to sufficient precision \citep{Savorgnano2026, Micheli2026}. In all of these systems, the grid's orientation is known only relative to the payload, so a separate measurement must place that frame on the sky, whether through ground-target photogrammetry, the attitude of the platform, or, as proposed for POLOCALC and CalSat, a star camera mechanically aligned to the polarizing filter \citep{Nati_2017, Johnson2015}. Despite the precision each of these steps can reach, combining them into the polarization angle received at the observatory remains a major source of complexity, especially for balloon- and space-based platforms where ground-target photogrammetry is not naturally available.

To address this challenge, we propose a calibrator architecture that measures the orientation of the polarizing element directly in the celestial frame, and with it the emitted polarization axis. Our approach is enabled by a single dual-use grid-optic, acting as both a millimeter-wave polarizer and an optical diffraction grating, which reduces reliance on separate payload attitude reconstruction and removes the need for a camera--grid alignment calibration. In this way, the grid itself is referenced to the sky in a single observation, so only the source--telescope line of sight must be supplied externally. Rather than replacing any particular artificial calibrator, our method is intended as a general framework that can be incorporated into any sky-source polarization calibrator.

Section~\ref{sec:concept} describes our method, shows how a single star camera image and the line of sight to the source determine a calibrator's polarization angle on the sky, and estimates how precisely. Section~\ref{sec:simvalidation} describes the simulations used to test our method, and Section~\ref{sec:results} what they show. In Section~\ref{sec:discussion} we discuss what these results imply, the requirements on a physical calibrator, and the path toward building and testing one.

\begin{figure*}[t]
\centering
\includegraphics[width=\textwidth]{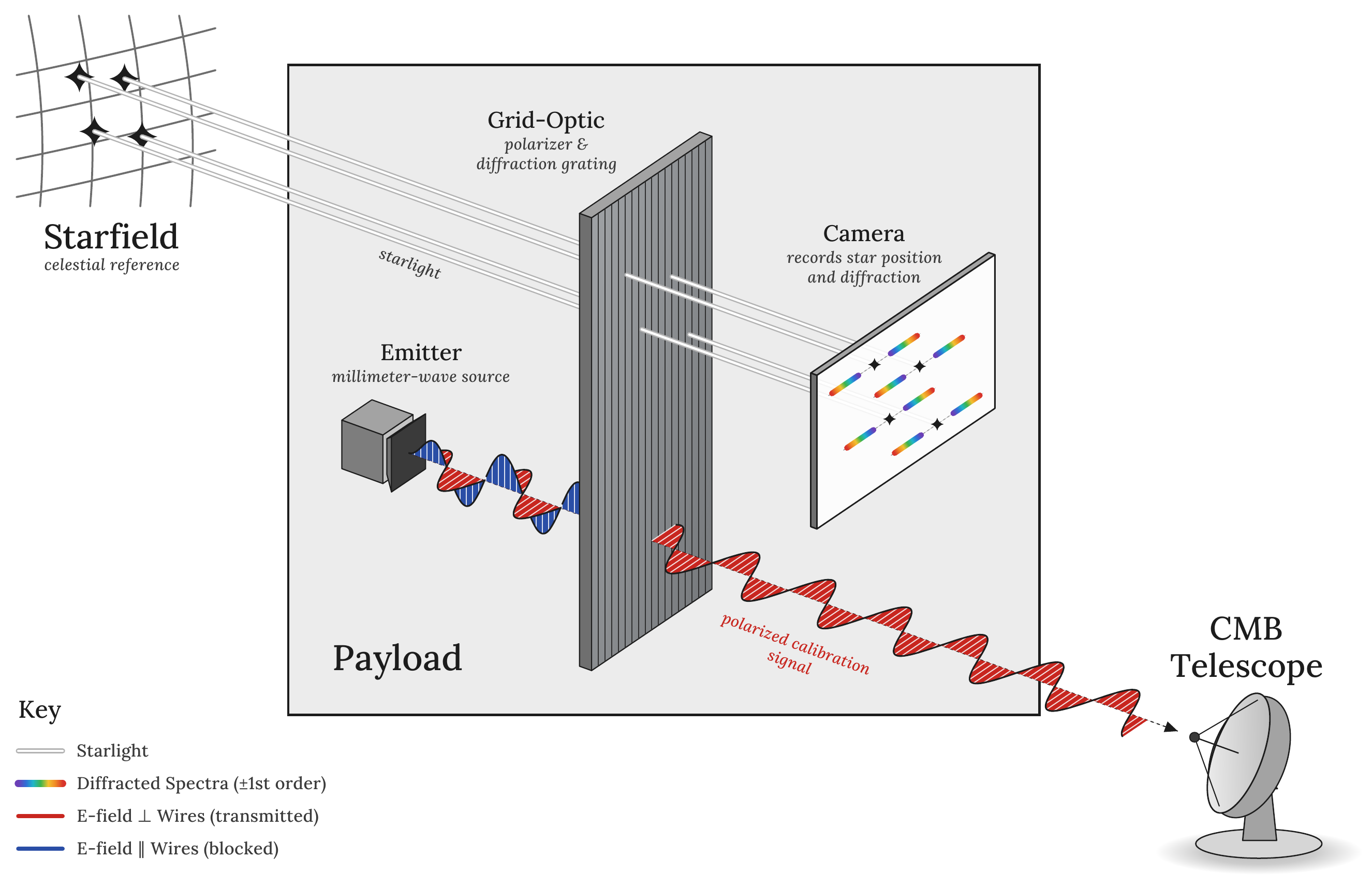}
\caption{A schematic of a calibrator using our method. A millimeter-wave emitter on the payload illuminates the grid-optic, which transmits only the polarization perpendicular to its wires toward the CMB telescope. Starlight passes through the same grid-optic into a star camera, which records each star together with its diffraction spectra (see Figure~\ref{fig:calibration_image} for an example image). The camera precisely measures the orientation of the diffraction pattern relative to the star field, and this gives the orientation of the grid on the sky. This in turn gives the orientation of the polarized calibration signal on the sky, which is what polarimeter calibration for CMB studies requires. A single grid-optic, which can be fabricated with established photolithography techniques, thus enables the angle calibration needed for $B$-mode searches. \vspace{.45cm}}
\label{fig:schematic}
\end{figure*}

\section{Calibration Method}
\label{sec:concept}

\subsection{Concept}
\label{sec:payload}

\begin{figure}[ht]
\centering
\includegraphics[width=0.95\columnwidth]{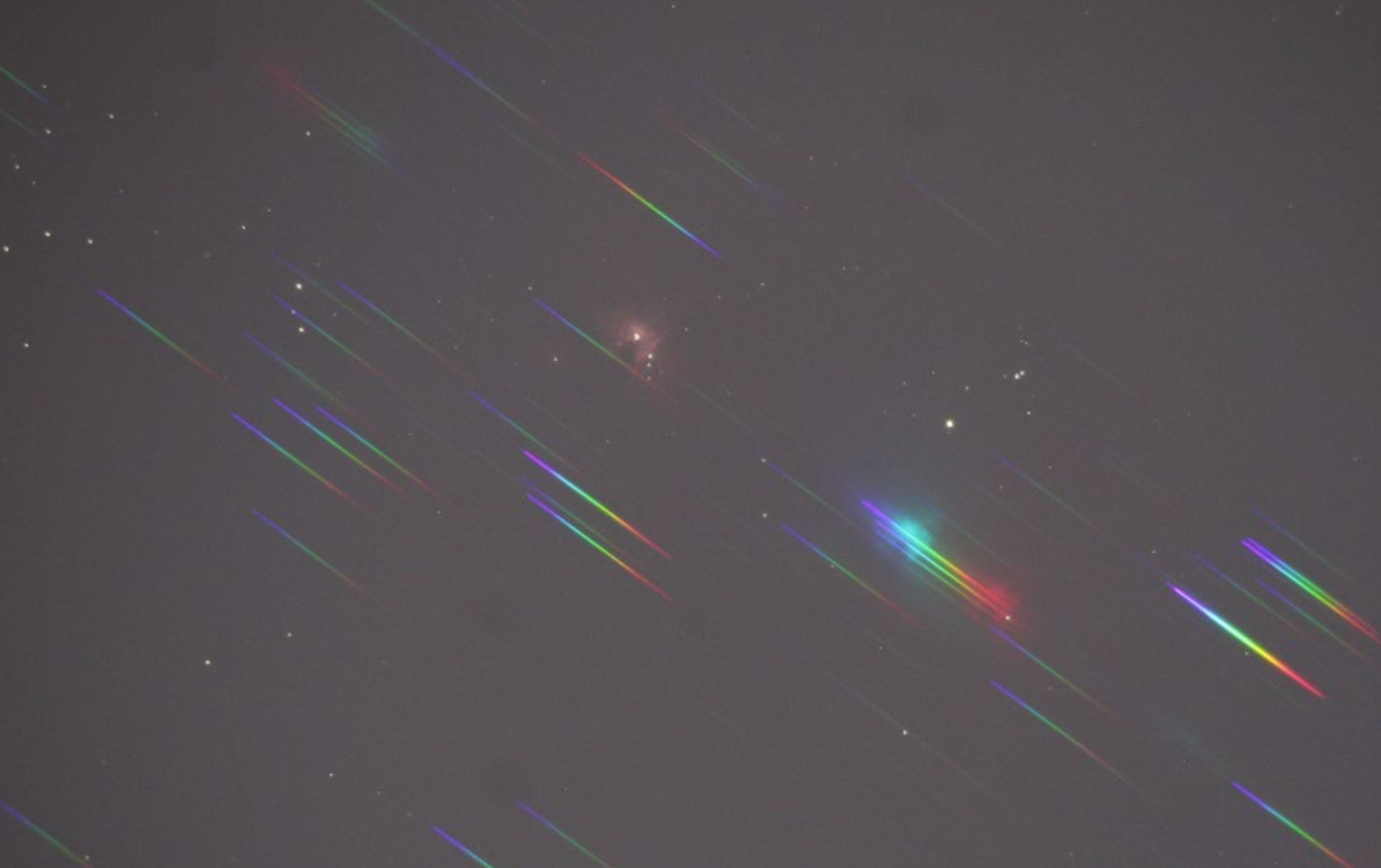}
\caption{A proof-of-concept star-field exposure, similar to the images our future calibration payload will produce, that we took with a telescope using a commercially available transmission diffraction grating. Alongside the point source stars are their dispersed first-order spectra, the optical signature that the grid-optic of Figure~\ref{fig:schematic} imprints on the star camera image. A preliminary analysis of this image gives the orientation of the RA/Dec grid from the star field, and the orientation of the diffraction grating with respect to that grid as expected and described further in Section~\ref{sec:next_steps}.
}
\label{fig:calibration_image}
\end{figure}

Our method takes the typical diffraction-based measurement of grid orientation demonstrated for systems such as PROTOCALC and COSMOCal and replaces the laser with starlight, so that the polarizing grid's orientation is obtained directly on the sky. We propose using a single periodic metallic grid-optic that polarizes a millimeter-wave calibration source and simultaneously diffracts visible starlight into a co-aligned star camera. A payload using our method, shown schematically in Figure~\ref{fig:schematic}, then carries three elements: a millimeter-wave source matched in frequency to the telescope being calibrated (e.g., 90\,GHz), a grid-optic, and a star camera. Starlight passes through the grid to the camera in the same direction that the calibration beam travels toward the telescope, so the camera sees the same sky behind the source as the telescope does. A single image from the star camera, like the example taken through an ordinary transmission grating in Figure~\ref{fig:calibration_image}, then contains both the star field, which ties the measurement to the sky, and the diffraction spectra of stars in the field, whose direction relative to those stars gives the grid's orientation, and therefore its polarization axis, on the sky.

This works because both the polarizing and diffracting functions of the grid-optic depend directly on the orientation of its wires. At visible wavelengths the grid period spans many wavelengths, so the grid acts as a diffraction grating that disperses light perpendicular to the wires \citep{Hecht2017,Goodman2017,Harvey1998}. At millimeter wavelengths the same period is much smaller than the wavelength, so the grid acts as a wire-grid polarizer that transmits the electric field perpendicular to the wires \citep{2001PASP..113..622H,Hecht2017}. Millimeter and visible wavelengths differ by a factor of several thousand, so one period can easily do both. A period of $10\,\mu$m, for example, is $200$--$330$ times smaller than the wavelengths of a $90$--$150$\,GHz beam, which favors a near-ideal polarizer compared with the larger periods commonly used in CMB instrumentation \citep{Matsuda2019}. It still diffracts $500$\,nm light by about $2.9^\circ$, far enough to separate each spectrum clearly from its star. Because one physical structure with a common wire direction sets both the geometry of the diffraction spectra and the polarization axis of the calibration beam, measuring the wire direction from the spectra also gives that polarization axis. We write this direction as $\hat{\mathbf{w}}=\hat{\mathbf{n}}\times\hat{\mathbf{g}}$, where $\hat{\mathbf{n}}$ is the grid-optic normal and $\hat{\mathbf{g}}$ is its grating vector, which lies in the grid plane and is perpendicular to the wires.

\subsection{Polarization Angle}
\label{sec:polangle}

To see how the wire direction determines the polarization a telescope receives, consider an ideal grid whose period is much smaller than the millimeter wavelength and whose wires transmit no field along their length. The field that gets through has no component along the wires, and it is perpendicular to its direction of travel at any angle of incidence \citep{Bird1960,2001PASP..113..622H}. The polarization axis received by a telescope must therefore be perpendicular to both the wires and the line of sight to the source. If $\hat{\mathbf{k}}_{\rm src\to tel}$ is the unit vector along this line of sight, from the calibration source to the telescope, the received axis is
\begin{equation}
\hat{\mathbf{p}}_{\rm tel}
=
\frac{\hat{\mathbf{k}}_{\rm src\to tel}\times\hat{\mathbf{w}}}
{\left|\hat{\mathbf{k}}_{\rm src\to tel}\times\hat{\mathbf{w}}\right|}.
\label{eq:p_tel}
\end{equation}
This is true wherever the telescope sits relative to the grid normal. It is also true however the emitter illuminates the grid, as long as all of the emitted radiation reaching the telescope has passed through the grid, because the same relation applies to each plane-wave component of the field leaving it \citep{Goodman2017}.

The telescope measures this polarization as an angle from north at the position where it sees the source on the sky. With $\hat{\mathbf{e}}_{\rm N,tel}$ and $\hat{\mathbf{e}}_{\rm E,tel}$ the north and east basis vectors the telescope uses at that position, the calibration angle, the polarization angle the telescope should measure, is
\begin{equation}
\psi_{\rm cal}
=
\operatorname{atan2}
\left(
\hat{\mathbf{p}}_{\rm tel}\cdot\hat{\mathbf{e}}_{\rm E,tel},
\hat{\mathbf{p}}_{\rm tel}\cdot\hat{\mathbf{e}}_{\rm N,tel}
\right),
\label{eq:psi_cal}
\end{equation}
measured from north through east (the IAU convention; in the convention common in CMB analyses, the angle is $-\psi_{\rm cal}$). The calibration angle is thus fixed by just two vectors: the wire direction, which the star camera image supplies, and the source--telescope line of sight, which is supplied externally. Because the camera measures $\hat{\mathbf{w}}$ against the sky, both vectors and the basis of Equation~\ref{eq:psi_cal} must be expressed in the celestial frame the telescope uses for its own polarization angles.

However, evaluating $\psi_{\rm cal}$ requires a specific telescope and its basis, which a star camera image alone does not have. To analyze the method at the level of the image, we therefore introduce a second angle, $\theta_{\rm sky}$. Because the star camera looks roughly along the continuation of the telescope--source line, its boresight $\hat{\mathbf{b}}$ (the sky direction of the image center) satisfies $\hat{\mathbf{b}}\approx-\hat{\mathbf{k}}_{\rm src\to tel}$, and Equation~\ref{eq:p_tel} gives $\hat{\mathbf{p}}_{\rm tel}$ parallel to $\hat{\mathbf{w}}\times\hat{\mathbf{b}}$. The calibration angle at the boresight is therefore the position angle of $\hat{\mathbf{w}}\times\hat{\mathbf{b}}$ there, measured east of north and defined modulo $180^\circ$. We call this the sky-frame grating orientation $\theta_{\rm sky}$. It is the calibration angle for a telescope looking exactly along the camera axis, and an image can state it by itself, with no specific source or observatory location. For an actual observation, the line of sight is offset from the camera axis by some angle $\gamma$, and $\psi_{\rm cal}$ then differs from $\theta_{\rm sky}$. Part of the difference is that north and east at the source's position differ from those at the boresight. The rest arises because the telescope sees the wires projected onto the plane perpendicular to its own line of sight. This projection turns them slightly with the offset alone, and any tilt of the wires toward or away from the camera axis, which we call their lean, adds a turn in proportion to the part of the offset that lies across the wires. Both differences are fixed by geometry and computed exactly by Equation~\ref{eq:psi_cal} from the full three-dimensional $\hat{\mathbf{w}}$ and the line of sight, so they add no uncertainty beyond that of these two vectors. This is one reason our method measures $\hat{\mathbf{w}}$ in three dimensions rather than only its orientation in the image.

\subsection{Geometry of the Diffraction Spectra}
\label{sec:conical}

A star camera image can determine $\hat{\mathbf{w}}$ because a grating displaces light only perpendicular to its wires, so a star and its diffracted light share the same direction component along the wires. To see this, let $\hat{\mathbf{d}}_\star$ be the direction from which a star's light arrives at the grid (its apparent direction on the sky), and $\hat{\mathbf{d}}_{m,\lambda}$ the apparent direction of its diffracted light in order $m$ at wavelength $\lambda$. For a grating of period $\Lambda$, the grating equation \citep{Harvey1998} acts on the components of these directions tangential to the grid,
\begin{equation}
\hat{\mathbf{d}}_{m,\lambda}-(\hat{\mathbf{d}}_{m,\lambda}\!\cdot\!\hat{\mathbf{n}})\,\hat{\mathbf{n}}
=
\hat{\mathbf{d}}_{\star}-(\hat{\mathbf{d}}_{\star}\!\cdot\!\hat{\mathbf{n}})\,\hat{\mathbf{n}}
+m\frac{\lambda}{\Lambda}\,\hat{\mathbf{g}} .
\label{eq:grating_vector}
\end{equation}
Since $\hat{\mathbf{w}}$ is perpendicular to both $\hat{\mathbf{n}}$ and $\hat{\mathbf{g}}$, taking the component along $\hat{\mathbf{w}}$ gives
\begin{equation}
\hat{\mathbf{d}}_{m,\lambda}\cdot\hat{\mathbf{w}}
=
\hat{\mathbf{d}}_{\star}\cdot\hat{\mathbf{w}}
\equiv c_\star .
\label{eq:cone}
\end{equation}
This holds at every order and wavelength, so a star and all of its diffracted light lie on a single cone of half-angle $\arccos c_\star$ about $\hat{\mathbf{w}}$. Equivalently, the tips of these unit vectors lie in a plane with normal $\hat{\mathbf{w}}$, so finding the wire direction amounts to finding the normal of that plane. A star's first-order spectrum is one arc of this cone, because light of wavelength $\lambda$ is displaced from the star by an angle $\approx\lambda/\Lambda$ along the grating vector. Each star has its own $c_\star$, but the wire direction is the same for all of them, so every spectrum in the field constrains the same $\hat{\mathbf{w}}$ and they can be combined to measure it.

This combined measurement must be done carefully, however, because a camera does not necessarily image spectra as straight parallel lines, though they may appear as such visually (e.g., Figure~\ref{fig:calibration_image}). A flat detector maps directions to positions by gnomonic projection, which images a cone as a straight line only when the cone opens out into a plane through the camera (i.e. $c_\star=0$). The spectra of all other stars are then curved arcs, each rotated by an amount that depends on its position in the field and grows with the square of the field size. A simple average of their orientations in the image would therefore miss the true wire direction by a few hundredths of a degree for a rectangular field several degrees across (the exact amount depends on the grating orientation), a large fraction of the $0.1^\circ$ goal. Our method instead builds directly on this geometry, determining the wire direction from the apparent directions on the sky, where Equation~\ref{eq:cone} holds exactly for a flat, periodic grid in front of the camera.

\subsection{Reconstructing the Wire Direction}
\label{sec:method}

A single star camera image contains everything needed to determine $\hat{\mathbf{w}}$ in this way. Its stars, matched to a catalog, give a plate solution---a World Coordinate System (WCS) that maps detector pixels to sky directions. Every point along a spectrum can then be mapped to the sky through the same solution, and there the spectra constrain $\hat{\mathbf{w}}$ in two ways. The direction each spectrum runs gives a first estimate (Section~\ref{sec:planefit}), and the position of each spectrum relative to the star that produced it, its parent star, gives a far more precise one, because star and spectrum are widely separated (Section~\ref{sec:pinned}).

\subsubsection{Spectrum-Direction Fit}
\label{sec:planefit}

By Equation~\ref{eq:cone}, the sky directions along each spectrum lie in a plane with normal $\hat{\mathbf{w}}$. Measured from its own mean direction, every spectrum therefore lies flat in a single common plane perpendicular to $\hat{\mathbf{w}}$, which makes the wire direction the one direction in which the spectra show the least scatter. To find it, each spectrum can be reduced to a series of centroids along its length, the intensity-weighted center of each cross-section. The wire direction is then the eigenvector with the smallest eigenvalue of the scatter matrix
\begin{equation}
\mathbf{S}
=
\sum_i W_i
\frac{\sum_j q_{ij}\,(\hat{\mathbf{d}}_{ij}-\bar{\mathbf{d}}_i)(\hat{\mathbf{d}}_{ij}-\bar{\mathbf{d}}_i)^{\mathsf T}}{s_i^2},
\label{eq:scatter}
\end{equation}
where $\hat{\mathbf{d}}_{ij}$ is the sky direction of centroid $j$ of spectrum $i$, $q_{ij}$ are the centroid weights, normalized to unit sum, $\bar{\mathbf{d}}_i$ is the weighted mean direction of the spectrum, and $s_i^2$ is its variance along its own length. Dividing by $s_i^2$ takes each spectrum's length out of the sum, so its contribution is set by its weight $W_i$ alone, which is chosen to favor long, narrow spectra because they define their directions best. The resulting estimate, $\hat{\mathbf{w}}_{\rm spec}$, is a full three-dimensional vector on the sky, so it requires no knowledge of the camera's alignment to the grid. Its precision is limited by how well each spectrum defines its own direction over its short length, so it serves mainly as the starting point for the star-anchored fit.

\subsubsection{Star-Anchored Fit}
\label{sec:pinned}

\begin{figure*}[t]
\centering
\includegraphics[width=\textwidth]{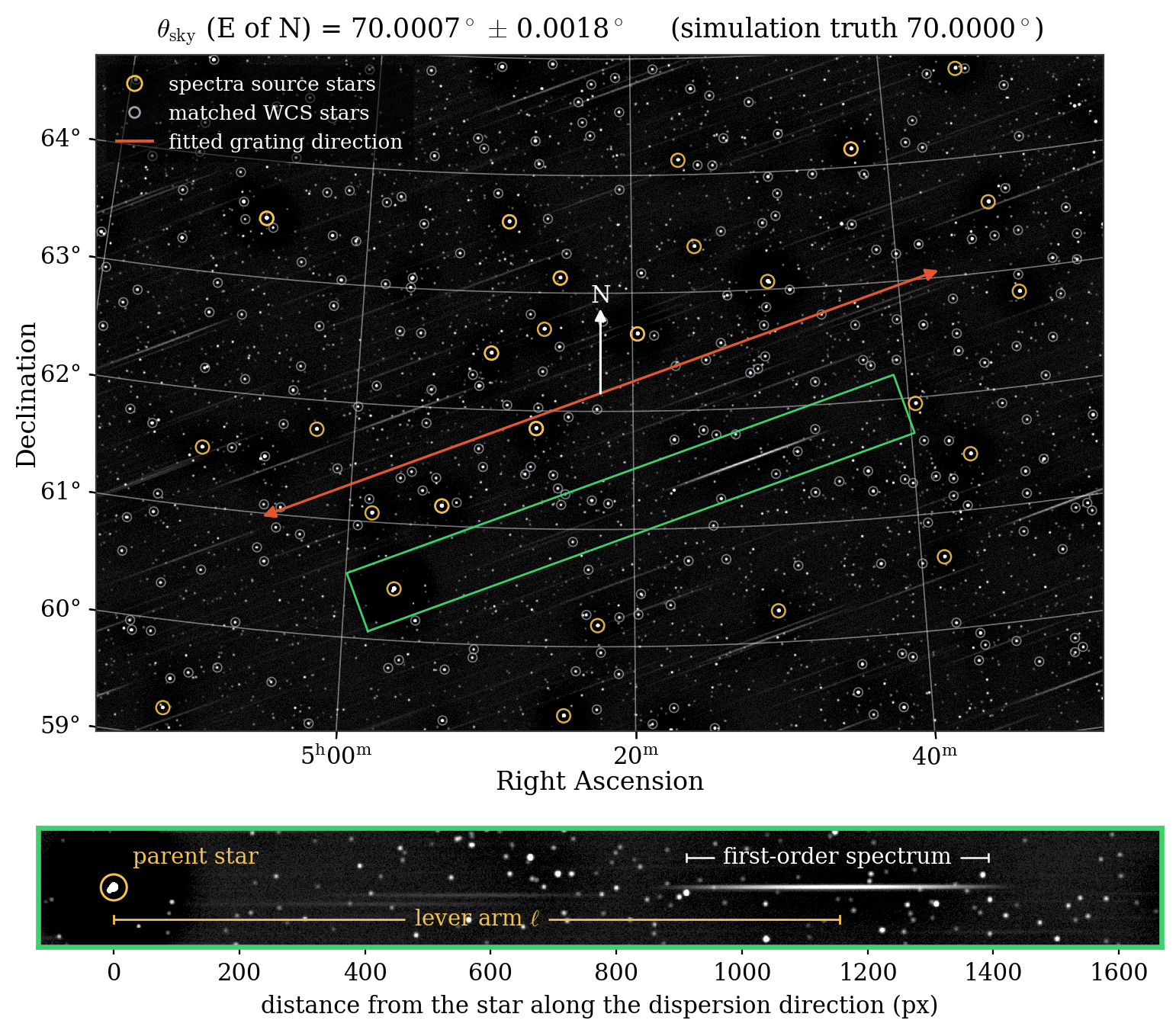}
\caption{
One simulated star camera image from the 1,000 used in this study (Section~\ref{sec:telesim}). The top panel shows the output of the analysis pipeline. Gold circles mark the parent stars of anchored spectra that lie on the sensor, and faint gray circles mark the other stars matched by the plate solution. The red arrow shows the fitted grating direction, perpendicular to the wires, and the white arrow points north at the field center, where $\theta_{\rm sky}$ is evaluated. The plate solution's right ascension and declination grid is overlaid. Our pipeline (Section~\ref{sec:extraction}) extracted a grid angle within $0.0007^\circ$ of simulation truth and a reported uncertainty of only $\pm0.0018^\circ$, well below the CMB goal of $0.1^\circ$ and in line with population-level values (Section~\ref{sec:results}). The bottom panel shows the region in the green box, resampled along the line from one parent star to its first-order spectrum. Here, it is possible to see more clearly the lever arm $\ell$ between the star and the centroid of its spectrum.
}
\label{fig:sim_result}
\end{figure*}

The star-anchored fit gains its precision from the distance between a star and its spectrum. A direction is measured far better between two well-separated points than along a short segment, and Equation~\ref{eq:cone} places each star on the cone of its own spectra, so a star and its spectrum together constrain $\hat{\mathbf{w}}$ over a much longer baseline than the spectrum alone. We call the separation between a star and the centroid of its spectrum the lever arm $\ell$ (bottom panel of Figure~\ref{fig:sim_result}). It is set by the first-order displacement at the spectrum's flux-weighted mean wavelength $\bar\lambda$, so for a camera of focal length $f$ and pixel pitch $p_{\rm pix}$ it is, in pixels,
\begin{equation}
\ell\simeq\frac{f\,\bar\lambda}{\Lambda\,p_{\rm pix}} .
\label{eq:lever}
\end{equation}
The centroid at the far end of this lever arm can be located across the spectrum's width to a small fraction of a pixel, $\sigma_c$, because it gathers the precision of the whole spectrum into one position. The great circle from the parent star to the centroid therefore fixes the orientation of $\hat{\mathbf{w}}$ to $\approx\sigma_c/\ell$, a much tighter constraint than the spectrum's own direction gives over its shorter length.

Using this constraint requires matching each spectrum to its parent star, which $\hat{\mathbf{w}}_{\rm spec}$ makes possible, since a parent star must lie on its spectrum's cone. Each matched, or anchored, spectrum then contributes its centroids, together with its parent star's direction, to the scatter matrix of Equation~\ref{eq:scatter}, with a weight $W_i=(\ell_i/\sigma_{c,i})^2$ so that spectra with long lever arms and well-measured centroids count the most. A better $\hat{\mathbf{w}}$ in turn improves the matching, so the two steps can be iterated, and the eigenvector with the smallest eigenvalue gives the final, star-anchored estimate of $\hat{\mathbf{w}}$. Like $\hat{\mathbf{w}}_{\rm spec}$, it is a full three-dimensional vector on the sky that requires no knowledge of the camera's alignment to the grid.

\subsection{Optical and Atmospheric Effects}
\label{sec:optical_effects}

In a real image, the camera's alignment and optics and the atmosphere disturb these measurements in several ways. Because our method relies on the position of each spectrum relative to its parent star on the sky, what matters most is whether a disturbance treats a star and its spectrum alike. Most do, so they are either absorbed by the plate solution or made irrelevant by the three-dimensional fit, and the rest can be corrected from the observing conditions or averaged out over the field.

This is because the camera and its plate solution together act as an instrument for measuring the direction of incoming light. The camera maps each direction on the sky to a position on the detector, and this map includes its pointing and orientation, the gnomonic projection, and any lens distortion. The stars, which pass straight through the grid in zeroth order and whose catalog directions are known, calibrate the whole map at once. Diffracted light passes through the same lens, so the plate solution also returns the direction of every point along a spectrum.

A tilt between the camera and the grid therefore does not affect the measurement, even though it can lean $\hat{\mathbf{w}}$ toward or away from the camera axis and changes the angle at which starlight meets the grid. The spectra then change shape and orientation across the field and shift slightly relative to their stars, so a fit that assumed the wires were perpendicular to the camera axis would be biased. On the sky, however, the plate solution has already accounted for the camera, and Equation~\ref{eq:cone} holds at any angle of incidence. Because $\hat{\mathbf{w}}$ is fit as a full three-dimensional vector, the lean is simply part of what is measured, so the camera need not be finely aligned to the grid (tested in Section~\ref{sec:sensitivity}).

Lens distortion is removed in the same way, provided the plate solution's distortion model, typically a polynomial such as the TAN-SIP form \citep{Shupe2005}, can represent the lens. A model that cannot leaves small errors in the sky directions of the spectra and therefore in $\hat{\mathbf{w}}$, but the image itself reveals this: stars miss their catalog positions, most visibly toward the edges of the field, and anchored spectra scatter more about their parent stars than expected, which also allows a better model to be chosen (Section~\ref{sec:extraction}).

Stellar aberration is also absorbed by this map. It displaces each star from its catalog direction before its light reaches the grating \citep{Kaplan2005}, but across a field a few degrees wide this is nearly a uniform shift, which the plate solution follows. Such a shift amounts to a small rotation about an axis perpendicular to the camera axis, so it only leans the fitted $\hat{\mathbf{w}}$ toward or away from the camera axis without turning it about that axis, and it therefore does not affect $\psi_{\rm cal}$ to first order.

Atmospheric refraction also displaces starlight before it reaches the grating, but it varies with zenith distance and wavelength, so it does have a small effect on the measurement. The plate solution is fit to refracted star positions and maps them to catalog directions, absorbing refraction as a smooth distortion. The grating, however, acts on light that has already been refracted, so the spectra are exact cones about $\hat{\mathbf{w}}$ only in the apparent directions, before the atmosphere's bending is removed. Mapped through the plate solution alone, they are deformed and the fitted $\hat{\mathbf{w}}$ rotates by up to about $0.02^\circ$ (for a drone observed from a high site at up to $60^\circ$ from the zenith), while for balloon- and space-based platforms, which lie above most of the atmosphere, the rotation is negligible.

This can be accounted for by carrying out both fits in the apparent frame, where Equation~\ref{eq:cone} is exact. The points along each spectrum are returned to this frame by refracting their plate-solution directions forward at the effective wavelength of the star images to which the solution was fit. The star-anchored fit needs one further step, because starlight of each wavelength reaches the grating from a slightly different direction. Each point is therefore compared with its parent star refracted forward at that point's own wavelength, which follows from its distance to the parent.

All of this requires only the zenith direction, from the observation time and site, and the atmospheric pressure and temperature, from weather data or a standard atmosphere at the camera's altitude. Any remaining error is then proportional to the fractional error of this refraction model, so measured weather data reduce it further.

Lateral chromatic aberration, or lateral color, is different again because it changes the map itself: the image scale of a lens varies slightly with wavelength, so different colors from the same direction land at slightly different distances from the image center. The plate solution is fit to broadband star images, so it cannot follow this, and spectrum centroids shift slightly relative to their parent stars. This shift is radial, however, pointing in opposite directions on opposite sides of the field, so it acts like a small change of scale, which cannot rotate the field as a whole. Its effect on the fitted orientation therefore largely averages out.

\subsection{Expected Precision}
\label{sec:error}

With these optical and atmospheric effects accounted for, the uncertainty on the calibration angle comes from the image itself, through the plate solution and the star-anchored fit, since the line of sight to the source is needed only coarsely (Section~\ref{sec:los}). To give each term a concrete scale, we evaluate it for a representative configuration typical of a small astronomy camera and lens: a $3096\times2080$-pixel sensor with $2.4\,\mu$m pixels behind a $50$\,mm, $f/2$ lens, observing in a $400$--$700$\,nm passband through a grid-optic with a period of $\Lambda=10\,\mu$m ($100\,{\rm lines\,mm^{-1}}$). This gives a field of view of about $8.5^\circ\times5.7^\circ$ at a scale of about $10''$ per pixel.

\subsubsection{Image-Level Uncertainty}
\label{sec:image_precision}

The first term is the orientation of the plate solution, which ties the image to the sky through its stars. A rotation error $\Delta\theta$ of the solution displaces a star at angular distance $r$ from the image center by about $r\,\Delta\theta$, so its orientation is best fixed by many precisely measured stars spread widely across the field. For $N$ matched stars with rms astrometric residual $\sigma_{\rm ast}$ and rms distance $r_{\rm rms}$ from the center, it is uncertain by
\begin{equation}
\sigma_{\rm WCS}
\sim
\frac{\sigma_{\rm ast}}
{\sqrt{N}\,r_{\rm rms}}.
\label{eq:wcs_orientation_scaling}
\end{equation}
For the representative configuration, about $300$ matched stars lie at $r_{\rm rms}\approx3^\circ$ from the center, and $\sigma_{\rm ast}\approx0.1$\,px ($1''$) is typical of centroiding and catalog errors for bright stars \citep{Lang2010}. This gives $\sigma_{\rm WCS}\approx0.0003^\circ$, more than two orders of magnitude below the $\sim0.1^\circ$ goal, so the stars fix the orientation of the image far more precisely than the calibration needs.

The second term comes from the spectra, and here the lever arm makes the difference. As a rough estimate, a typical usable spectrum defines its own direction to about $0.1^\circ$, so some $50$ such spectra would determine $\hat{\mathbf{w}}_{\rm spec}$ to $0.1^\circ/\sqrt{50}\approx0.014^\circ$. In the star-anchored fit, each spectrum instead constrains the orientation over the lever arm, so $K_{\rm anc}$ anchored spectra give
\begin{equation}
\sigma_{\rm diff}
\sim
\frac{\sigma_c}{\ell\,\sqrt{K_{\rm anc}}} .
\label{eq:anchored_design}
\end{equation}
At $\bar\lambda=550$\,nm, the representative configuration has a lever arm of $\ell\simeq1150$\,px, so with a centroid uncertainty of a quarter of a pixel and $K_{\rm anc}=50$, $\sigma_{\rm diff}\approx0.0018^\circ$, nearly an order of magnitude below the spectrum-direction fit.

Both terms can be evaluated from the image itself, $\sigma_{\rm WCS}$ from the residuals and positions of the matched stars, and $\sigma_{\rm diff}$ from the curvature of the weighted star-anchored fit (how quickly its residuals grow as $\hat{\mathbf{w}}$ moves away from its best value). A single image can therefore report its own uncertainty on $\theta_{\rm sky}$,
\begin{equation}
\sigma_{\rm sky}=\left(\sigma_{\rm WCS}^2+\sigma_{\rm diff}^2\right)^{1/2}.
\label{eq:sigma_sky}
\end{equation}

This statistical uncertainty does not include systematic errors, such as a spectrum matched to the wrong star or a plate solution whose distortion model does not suit the lens (Section~\ref{sec:optical_effects}). However, because an image constrains $\hat{\mathbf{w}}$ in several ways at once, such errors show up in the image itself as disagreements: anchored spectra scatter about their parent stars more than their centroid uncertainties predict, and the estimate changes between the spectrum-direction and star-anchored fits or between distortion models. A reported uncertainty can therefore be widened whenever such a disagreement is significant, so that it continues to describe the actual error.

The uncertainties above concern the orientation of $\hat{\mathbf{w}}$ about the camera axis, which sets the uncertainty on $\theta_{\rm sky}$. The fit measures the lean of $\hat{\mathbf{w}}$ as well (Section~\ref{sec:polangle}), but less precisely, because a lean leaves the overall direction of the spectra unchanged and shows only through perspective, as a gradual turning of their directions that grows with distance from the image center along the direction of dispersion. Its uncertainty is therefore larger than $\sigma_{\rm diff}$ by roughly the inverse of the field's angular extent, $\sigma_{\rm out}\sim\sigma_{\rm diff}/r_{\rm rms}$, with $r_{\rm rms}$ the rms angular distance from the image center in radians, whatever the size of the lean itself.

Whether this uncertainty reaches $\psi_{\rm cal}$ depends on $\gamma$. Because the camera sees the sky behind the source as the telescope does (Section~\ref{sec:payload}), $\gamma$ is simply how far from the image center the telescope's line of sight falls, and it is known for each exposure from the plate solution and the telescope's pointing. The lean enters $\psi_{\rm cal}$ only through the part of this offset that lies across the wires, so $\sigma_{\rm out}$ has no effect on $\theta_{\rm sky}$. Since the lean itself is accounted for exactly, only its uncertainty matters, and it adds at most $\sigma_{\rm out}\sin\gamma\sim\sigma_{\rm diff}\,\gamma/r_{\rm rms}$ to the uncertainty on $\psi_{\rm cal}$. This term vanishes when the line of sight falls at the image center and reaches the size of the image-level uncertainty only toward the edge of the field, so a calibrator whose camera is aligned with its beam, and whose beam is pointed at the telescope, keeps it negligible. It can also be evaluated for each exposure, because the fit gives the full uncertainty of $\hat{\mathbf{w}}$, including the correlation between orientation and lean that appears when most anchored spectra lie on the same side of their parents. Propagating it through Equation~\ref{eq:p_tel} at that exposure's line of sight then gives the uncertainty on $\psi_{\rm cal}$.

\subsubsection{Line of Sight}
\label{sec:los}

The other vector that fixes the calibration angle, the line of sight to the source, needs to be known only coarsely, and the telescope's own pointing can supply it. A small error in the line of sight could change the angle in two ways. The first is through the computed polarization axis of Equation~\ref{eq:p_tel}. The wires lie nearly perpendicular to the line of sight, so an error in it moves $\hat{\mathbf{p}}_{\rm tel}$ almost entirely along the line of sight itself, and a telescope cannot detect polarization along that direction.

The second is through north and east, whose directions differ from one place on the sky to another, especially near the celestial pole. An error in the line of sight is also an error in the source's position on the sky, so it rotates the basis of Equation~\ref{eq:psi_cal}. This does not matter as long as the predicted and measured angles use the same position. The telescope measures its angle at the position where it sees the source, even if that position is slightly wrong, so evaluating the predicted angle at that same position rotates both together and the rotation cancels. Only a prediction made at a different position, such as one from GNSS, would carry this rotation into $\psi_{\rm cal}$.

What remains is the line-of-sight error times the small angle by which the wires depart from perpendicular to the line of sight, an angle set by $\gamma$ and the lean, and this is negligible for a grid that faces the telescope. Refraction of the beam on its way to the telescope bends it within a vertical plane, so it acts only as a small line-of-sight error and falls under the same argument. The line of sight therefore only has to be good enough to set $\hat{\mathbf{k}}_{\rm src\to tel}$ in Equation~\ref{eq:p_tel} and the offset $\gamma$.

\section{Simulation Methods}
\label{sec:simvalidation}

Overall, our method depends on recovering both the diffraction spectra and the plate solution from a single image. The physics of Section~\ref{sec:concept} sets the precision this should allow, but reaching it in practice requires software that can extract these quantities from a realistic image, with its crowded stars, overlapping spectra, and noise, as well as misalignment between the camera and grid-optic and the other effects of Section~\ref{sec:optical_effects}. We therefore test the full measurement on images with known ground truth, using two separate software packages. An image simulator renders a synthetic exposure from a physical description of the sky, atmosphere, optics, and detector, and an angle-extraction pipeline measures the wire direction from an image (which sets the emitted polarization axis; Section~\ref{sec:polangle}). To keep the test independent of the method it checks, the simulator renders every spectrum from the full vector grating equation (Equation~\ref{eq:grating_vector}), while the pipeline's fits rely only on its consequence that each star's spectra lie on a common cone. Neither package calls the other, and where both model an effect, each has its own implementation.

\subsection{Simulated Images}
\label{sec:telesim}

We built a Python package, tele-img-sim \citep{bass2026tele_img_sim}, to generate synthetic star camera images for this project. Star fields are drawn from Gaia Data Release 3 down to $G=12$ mag, so stellar positions, magnitudes, colors, and densities reflect real patches of sky \citep{GaiaDR3}. Each star is given a blackbody spectrum with a temperature derived from its Gaia color, and its first-order spectra are projected onto the detector, wavelength by wavelength, through models of the lens and camera. The grid-optic is modeled as covering the lens's entrance pupil, and because a spectrum lands a full lever arm from its star, stars beyond the edge of the sensor are included wherever their spectra fall on it. The image is then degraded by sky background, atmospheric seeing, pointing jitter, and photon and read noise. For each image, the simulator records the exact three-dimensional wire direction and an exact astrometric solution, which give unambiguous ground truth.

All simulations use the representative configuration of Section~\ref{sec:error}. We ran 1000 fields, each with a random pointing and random orientations of both the camera and the grid-optic, so that the wires can lie at any angle on the detector, and with a camera--grid tilt of up to $2^\circ$ about a random axis. Each field also draws its own atmospheric and optical effects at random. Starlight is refracted before it reaches the grating, at a $5300$\,m site with the field center $20$--$60^\circ$ from the zenith, under a pressure, temperature, and humidity that depart from the standard dry atmosphere by up to $3\%$, $20$\,K, and $80\%$ relative humidity. The lens then adds radial distortion of either sign, with a third-order term reaching $1$--$3\%$ at the sensor corner and a fifth-order term of up to half that size, and lateral color of up to $\pm1$\,ppm\,nm$^{-1}$.

\subsection{Angle Extraction}
\label{sec:extraction}

We developed SpectrAngle \citep{bass2026spectrangle}, a Python pipeline that implements the reconstruction of Section~\ref{sec:method}. It takes a single image, simulated or real, and returns the wire direction with its uncertainty $\sigma_{\rm sky}$, together with a summary figure like the top panel of Figure~\ref{fig:sim_result}. The image is first plate solved, since every later step works in sky directions. Point sources are detected with the elongated spectra masked, so that the spectra are not mistaken for stars, and matched with \texttt{solve-field} \citep{Lang2010} against an index built from Gaia DR3, the same catalog from which parent stars are later drawn. Because the right distortion model for a lens is not known in advance, each image is solved with both third- and fifth-order TAN-SIP polynomials, and the rest of the reconstruction is carried out under each. For each solution, $\sigma_{\rm WCS}$ is computed from the matched stars after excluding the few whose residuals lie far beyond the rest, so that occasional mismatches cannot inflate it.

The spectra are then found as elongated features, cleared of overlapping starlight, and reduced to centroids along their length, and any whose direction departs strongly from the common plane of the spectrum-direction fit is rejected. Matching the remaining spectra to their parent stars is the step most prone to error, because many spectra have more than one star near their cone. Each candidate is therefore scored on its distance from the cone, on whether its brightness matches the spectrum's flux, and on whether a spectrum of the opposite order points back to it, and a spectrum is anchored only when its best candidate is decisively preferred. The pipeline then adopts the plate solution under which the anchored spectra scatter least about their parent stars, since a poor distortion model shows up as exactly this scatter. Finally, the reported uncertainty is scaled up if the anchored spectra scatter more than their centroid uncertainties predict, and widened if the two fits or the two plate solutions disagree by more than three times their uncertainty.

The pipeline is given only what an observer would have: the image, the Gaia catalog with colors, the nominal instrument parameters, the zenith direction, and a standard atmosphere at the site altitude. It receives no simulator truth, such as the orientations of the camera and grid-optic, the lens distortion, or the drawn weather, so it handles a simulated image exactly as it would a real one.

\section{Results}
\label{sec:results}

A single simulated exposure determines the wire direction to a few thousandths of a degree, the precision estimated in Section~\ref{sec:error}, and the uncertainty reported from each image alone is slightly conservative, so it can be trusted where no ground truth exists.

\subsection{Accuracy and Reported Uncertainty}
\label{sec:sim_results}

We distinguish the error of the reconstruction, the angle between the recovered and true orientation of the grid on the sky, from the uncertainty the reconstruction reports from the image alone. The error measures accuracy but is known only in simulation, whereas the reported uncertainty is what a real observation provides, so it must capture the size of the error without access to ground truth.

\begin{figure*}[t]
\centering
\includegraphics[width=\textwidth]{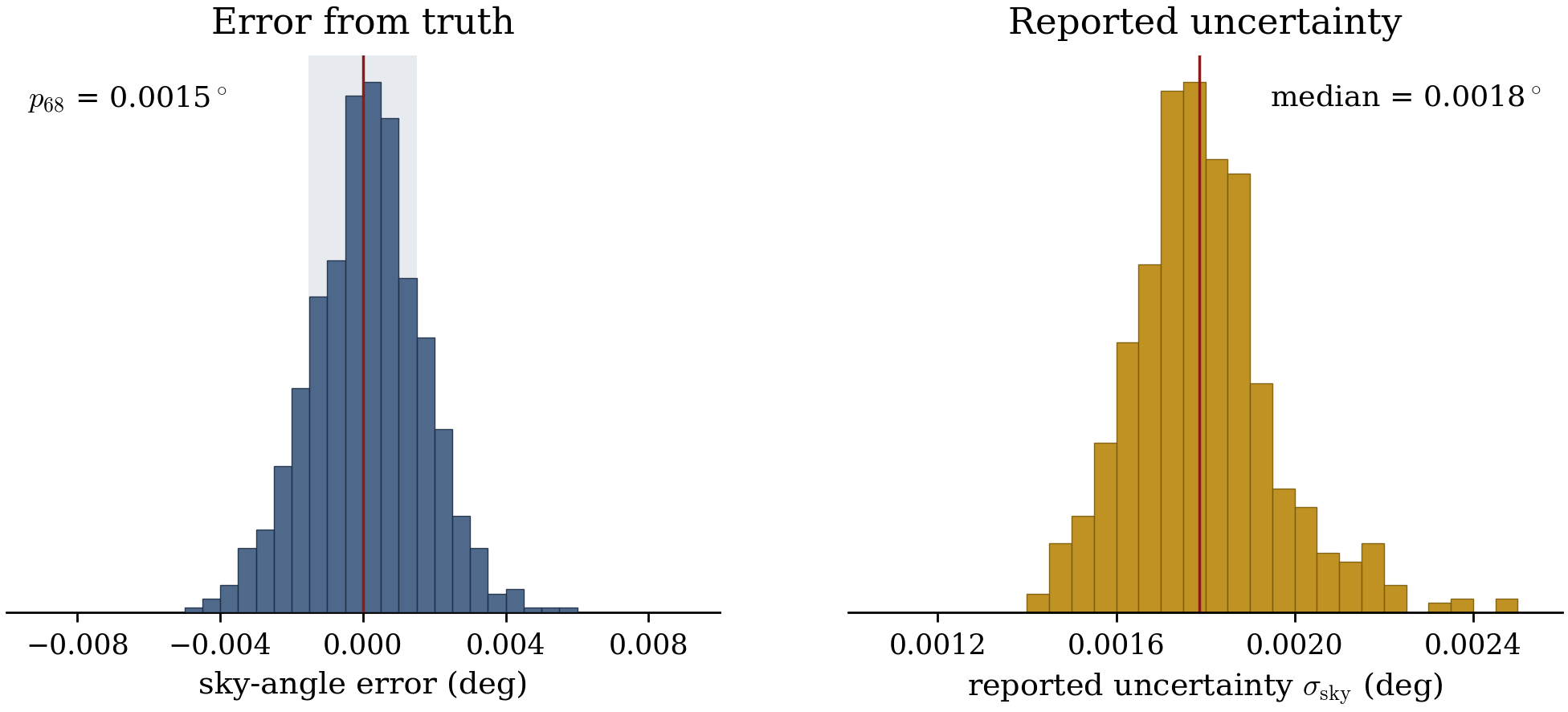}
\caption{
Histograms showing our pipeline's angle analysis results from all 1,000 simulated star field images. The left panel shows the recovered angle minus the simulation truth. The 68th percentile of the error from truth (shaded gray) is only $\pm 0.0015^\circ$, showing that our method recovers injected grid-optic angles on the sky extremely well. The right panel shows the reported uncertainty $\sigma_{\mathrm{sky}}$ for each field. For these measurements the median reported uncertainty---which is the value that matters for real data, computed with no knowledge of the ground truth---is just $0.0018^\circ$. Together these plots show that, in principle, single images from our method contain all the information required to constrain a calibrator's emitted polarization orientation to well below the $0.1^\circ$ level.
}
\label{fig:uncertainty_histograms}
\end{figure*}

Figure~\ref{fig:uncertainty_histograms} shows that both behave well across the simulated fields. The errors form a narrow distribution centered on zero, with a 68th percentile of their absolute value of $p_{68}=0.0015^\circ$, about ten times below that of the spectrum-direction fit alone ($0.014^\circ$), and 99\% of fields lie within $0.005^\circ$ of the truth. The reported uncertainties are just as consistent from field to field, with a median of $\sigma_{\rm sky}=0.0018^\circ$, matching the estimate of Section~\ref{sec:error}. Most importantly, they are not overconfident. When each field's error is compared with its own reported uncertainty, $77\%$ of errors lie within $\sigma_{\rm sky}$ and $97\%$ within $2\sigma_{\rm sky}$, more than the $68\%$ and $95\%$ expected for a Gaussian. Even the rare large errors are recognizable, because all five fields with errors above $0.01^\circ$, including the single one beyond the $0.1^\circ$ goal, report unusually large uncertainties, and four of them were widened by the checks of Section~\ref{sec:extraction}.

\subsection{Sensitivity to Optical and Atmospheric Effects}
\label{sec:sensitivity}

The modeled effects, including the misalignment between the camera and grid-optic, add only slightly to the error, and none leaves a trace that grows with its size. Because every field draws its own combination of effects, such a trace would appear as a correlation between the error and one of the drawn parameters. Neither the signed nor the absolute error correlates significantly with any of them: the camera--grid tilt and its axis, the zenith distance and direction, the departure of the weather from the standard atmosphere, lens distortion, or lateral color. The largest Spearman coefficient among these tests is $|\rho|=0.07$, consistent with chance for this many tests. To measure each contribution directly, we also rendered a subset of the fields with each effect alone and with none, and compared the typical errors on the same fields. Measured this way, no modeled effect contributes more than about $0.001^\circ$ in quadrature, below the statistical uncertainty of a single image.

\subsection{Uncertainty Budget}
\label{sec:error_summary}

The two image-level terms of Section~\ref{sec:error} agree with what the simulated images report: the plate-solution term, estimated at $0.0003^\circ$ for the representative configuration (Equation~\ref{eq:wcs_orientation_scaling}), has a median reported value of $0.0002^\circ$ across the simulated fields, and the star-anchored term, estimated at $0.0018^\circ$ (Equation~\ref{eq:anchored_design}), has a median of $0.0018^\circ$. Combined in quadrature, they give $\sigma_{\rm sky}\simeq0.002^\circ$ (Equation~\ref{eq:sigma_sky}), the uncertainty on $\theta_{\rm sky}$ and on the calibration angle for a line of sight near the center of the camera's field. For other lines of sight, the lean of $\hat{\mathbf{w}}$ is accounted for exactly, and its uncertainty adds a term that remains negligible across the central part of the field (Section~\ref{sec:image_precision}), so we do not list it separately. The simulations bear out the scaling behind this: the fitted lean is uncertain by about $\sigma_{\rm diff}/r_{\rm rms}$, independent of the size of the simulated camera--grid tilt. An error in the line of sight itself, by contrast, adds almost nothing (Section~\ref{sec:los}). These terms describe the measurement itself, and Section~\ref{sec:tolerances} discusses the hardware tolerances of a physical calibrator.

\section{Discussion}
\label{sec:discussion}

\subsection{Implications for Polarization-Angle Calibration}
\label{sec:implications}

Our method measures the orientation of the grid that sets the calibration beam's polarization directly in the frame in which the calibration is needed. A single image therefore replaces the chain of laboratory alignment, attitude reconstruction, and coordinate transformations that otherwise links the emitted polarization axis to the sky, and the alignment between the camera and grid-optic need not be known. Each exposure yields the wire direction, averaged over the exposure, together with its uncertainty, independently of any other exposure, so in operation a time-stamped sequence of exposures samples $\psi_{\rm cal}$ throughout an observation of the source. The measurement needs a night sky with enough stars, and its precision is set by the lever arm (fixed by the focal length and grating period), the centroid precision, and the number of usable spectra, so it can be estimated before hardware is built, as with our simulations.

For calibration, the image-level uncertainty of $\sim0.002^\circ$ lies fifty times below the $\lesssim0.1^\circ$ goal of current and next-generation experiments such as the Simons Observatory \citep{Bryan2018, abitbol_2021}. It is also well below the $0.045^\circ$ in-flight roll accuracy reported for PROTOCALC \citep{Coppi2025} and the $\sim 0.03^\circ$ internal precision COSMOCal reached at the IRAM 30\,m telescope \citep{Savorgnano2026}, although these comparisons set a simulated result against real measurements. A reference at the $\sim0.002^\circ$ level would leave substantial margin for the other terms in a telescope's polarization-angle budget, including for tests of cosmic birefringence. It is also independent of the $EB$ self-calibration, whose degeneracy with birefringence motivates external calibrators in the first place \citep{Keating2013, Minami2019}.

Measuring the grid directly against the sky also means that neither the payload's attitude nor its position needs to be known precisely, because the position enters only through the line of sight, which the telescope's own pointing can supply (Section~\ref{sec:los}), and the attitude enters only coarsely, through where that line of sight falls in the camera's image (Section~\ref{sec:image_precision}).

\subsection{Scope of the Simulations}
\label{sec:scope}

These results are for a single representative configuration. A different lens, sensor, or grating changes the lever arm, the number of usable spectra, and the centroid precision, and with them the numbers reported here, but the method and the form of its uncertainty estimate carry over.

Real observations will also be less ideal than the simulated ones. In the simulations, the pipeline shares the simulator's star catalog, instrument parameters, and refraction formula, the lens distortion is radial, and stellar aberration, which the plate solution absorbs, is not simulated, nor is rotation of the payload during an exposure. A real instrument adds fainter, uncatalogued stars, imperfectly known instrument parameters, more general distortion, and payload motion, as well as the hardware tolerances discussed below.

\subsection{Requirements on a Physical Calibrator}
\label{sec:implement}

For a physical calibrator using our method, the practical questions are then whether its grid-optic diffracts enough starlight for the camera to detect, whether it can be fabricated, and how small the imperfections of its hardware must be to leave the measured angle unbiased.

\subsubsection{Optical Throughput and Detectability}
\label{sec:throughput}

A wire-grid polarizer acts as an ordinary grating at visible wavelengths, so the question is only how much of the starlight it diffracts into the first orders. If the grid is treated as a binary (opaque/transparent) grating whose open, transmitting fraction is $a$, the fraction of the incident intensity diffracted into order $m$ is approximately
\begin{equation}
\eta_m \simeq \left[a\,\mathrm{sinc}(ma)\right]^2,
\end{equation}
with $\mathrm{sinc}(x)=\sin(\pi x)/(\pi x)$ \citep{Hecht2017, Goodman2017}. For $a\simeq0.5$, typical of wire-grid polarizers, this gives $\eta_0\simeq25\%$ in the undiffracted zeroth order and $\eta_{\pm1}\simeq10\%$ in each first order, while the higher orders are much weaker, since the even orders vanish and the third carries only about $1\%$.

With about $10\%$ in each first order, the brighter stars already give enough detectable spectra for the measurement. At the same time, the zeroth-order star field must stay unsaturated while the spectra stay above the noise floor, so detectability becomes a design constraint on the grating period, duty cycle, lens f-ratio, passband, and exposure, which tele-img-sim (Section~\ref{sec:telesim}) can be used to explore for a given platform. In practice, a real grid, with finite wire thickness and a substrate, will differ somewhat from this simple estimate, but its efficiency is straightforward to measure directly by illuminating a millimeter-wave polarizer with visible light.

\subsubsection{Fabrication Feasibility}
\label{sec:fabrication}

Such a grid is also well within current fabrication practice. Wire-grid polarizers with sub-wavelength periods, from the ultraviolet through the terahertz, are routinely made by photolithography, nanoimprinting, and related techniques that pattern metal on dielectric or semiconducting substrates \citep{Chen2025}. Free-standing metallic wire grids at the coarser periods of millimeter-wave polarimetry have likewise been built and characterized for astrophysical instruments, including the grids of variable-delay polarization modulators for CMB polarization measurements \citep{Voellmer2008}. At a period of $10\,\mu$m and a duty cycle near one half, the wires are a few micrometers wide, well within the reach of lithographic patterning, and they need only be thick compared with the skin depth at millimeter wavelengths. Building a grid-optic for our method therefore comes down to choosing its period, duty cycle, metal thickness and conductivity, and substrate so that the one grid both polarizes the beam well and diffracts enough starlight, and both properties can then be checked directly on a fabricated grid by measuring its millimeter-wave extinction ratio and its optical diffraction efficiency.

\subsubsection{Hardware Tolerances}
\label{sec:tolerances}

Even in a well-made calibrator, small imperfections in the hardware can bias the measured angle in two ways. They can make the wires the camera measures differ from those that set the beam's polarization, or they can make the camera image a star and its spectrum differently. The first arises most directly because the star camera and the calibration beam pass through different regions of the grid-optic, so any change in the direction of the wires between those regions enters $\psi_{\rm cal}$ one-to-one. To stay below the image-level uncertainty, the wire direction must therefore be uniform to $\lesssim0.002^\circ$ across the parts of the grid used by both, or about $3.5\,\mu$m over $100$\,mm. Tilts of the grid surface between those regions instead give the wires a slightly different lean in each, which, like any lean (Section~\ref{sec:polangle}), enters $\psi_{\rm cal}$ only in proportion to the offset of the line of sight, and so only at second order when the telescope views the grid near its normal.

Even a perfectly uniform grid does not set the polarization entirely by itself, because a real grid also lets through a small part of the field along its wires, which Equation~\ref{eq:p_tel} neglects. For an unpolarized source this leaked field only slightly lowers the degree of polarization, without changing its angle. Bright millimeter-wave sources, however, are usually coherent and linearly polarized, and if their polarization is not perpendicular to the wires, the leaked field can pull the transmitted polarization slightly toward it. This pull is only a small fraction of the emitter's misalignment, so the emitter needs far less precise alignment to the grid than the grid's own orientation on the sky. Likewise, the wires should be the last element in the millimeter-wave path, facing the telescope, since a dielectric layer after them could rotate the polarization slightly at oblique incidence.

Imperfections in the camera's optics act instead through the star-anchored fit, which compares the position of each star with that of its spectrum. The plate solution absorbs any shift that stars and their spectra share, but not a difference between them. Such a difference can arise because a star's image mixes all colors, while each point along a spectrum holds only a narrow band of color, so aberrations that vary with color, such as chromatic coma, can center the two slightly differently. For a well-centered lens these shifts are symmetric about the image center, so, like lateral color, they largely average out. What remains is the small part that pushes the spectra consistently to one side across the field. This part, $\Delta$, biases the orientation by about $\Delta/\ell$, so keeping it below the image-level uncertainty requires $\Delta\lesssim0.03$\,px. This is tight, but it applies only to the asymmetric remainder and can be checked in the laboratory. Likewise, any window or substrate in the starlight path, including the grid's own, should have faces parallel to good optical precision, since a slight wedge between them acts as a weak prism that spreads every star's light by color in the same direction. Beyond these, only smaller terms remain, from the detector, such as pixel-to-pixel variations in sensitivity, and from departures of the real atmosphere from the refraction model, while the Gaia DR3 positions and proper motions of the parent stars \citep{GaiaDR3} are accurate far below the level that matters here.

Overall, none of these requirements is unusual for a well-built optical instrument, and those on the hardware can each be checked in the laboratory before flight.

\subsection{Toward a Working Calibrator}
\label{sec:next_steps}

Initial tests with an off-the-shelf star camera and lens, using an ordinary transmission grating in place of the grid-optic, already show our method working as expected. With SpectrAngle, we processed night-sky images from this hardware end-to-end, including the one in Figure~\ref{fig:calibration_image}, plate solving each and detecting and measuring the stars and their spectra together in real, non-ideal exposures. As in the simulated images, the direction of the spectra changes across the field in the way the conical geometry of Section~\ref{sec:conical} predicts, and for the image in Figure~\ref{fig:calibration_image} the reported uncertainty is consistent with what our estimates in Section~\ref{sec:error} predict for that hardware.

The next step, already under way, is to obtain or fabricate a dual-use grid-optic and test it in the same way, and then to build an integrated payload that carries it together with the star camera and millimeter-wave source. We will fly this payload on a drone as an initial test bed, measuring its signal from the ground to verify the full chain from star camera image to received polarization angle outside the laboratory. We will then observe it with a CMB telescope, comparing $\psi_{\rm cal}$ directly against the telescope's own polarization-angle response. Throughout, we plan to work in conjunction with other calibration efforts, so that our method can be checked against, and combined with, existing approaches. Beyond this initial test bed, nothing in our method is specific to drones, so the same approach extends to balloon- and space-based calibrators.

\section{Conclusions}
\label{sec:conclusions}

Artificial polarized sources can calibrate the absolute polarization angle of CMB telescopes only if their emitted polarization angle is known in the celestial frame, and transferring that angle from the payload frame to the sky has been a central limitation. We have proposed a calibrator architecture in which a single dual-use grid-optic both polarizes the calibration beam and diffracts starlight into a star camera, so that the wire direction, and with it the emitted polarization axis, is measured directly on the sky. The star-anchored fit makes this measurement precise by using the long lever arm between each star and its first-order spectrum. In simulations of 1000 real star fields, a single exposure determines the grid's orientation in the celestial frame with an uncertainty of about $0.002^\circ$, fifty times below the $0.1^\circ$ goal, and each image's own reported uncertainty is slightly conservative, so it can be trusted where no ground truth is available. Because the grid is measured directly against the sky, neither the payload's attitude nor its position needs to be known precisely.

Our method thus turns the measurement of a calibrator's polarization axis on the sky into a single astrometric observation, for drone-, balloon-, and space-based sources alike. Beyond these simulations, initial tests on real images already show it working as expected, and future work will check the hardware tolerances of Section~\ref{sec:tolerances}. Ultimately, our calibration method, built on a single dual-use grid-optic, could give CMB experiments an absolute polarization-angle reference at the few-thousandths-of-a-degree level from every exposure, supporting searches for primordial gravitational waves and independent tests of cosmic birefringence.\\

\begin{acknowledgments}

%
%
We are grateful to Declan Baker and Patricio Gallardo from the Devlin Lab at the University of Pennsylvania.
This work also benefited from useful conversations with Federico Astori, Anna Fox-Ryan, and Annika Deutsch about the nature of the light in the universe.

\end{acknowledgments}

\software{SpectrAngle \citep{bass2026spectrangle},
          tele-img-sim \citep{bass2026tele_img_sim},
          Astrometry.net \citep{Lang2010},
          Astropy \citep{astropy:2013, astropy:2018, astropy:2022},
          NumPy \citep{numpy2020},
          SciPy \citep{scipy2020},
          Matplotlib \citep{matplotlib2007},
          Photutils \citep{photutils2026}}

\bibliography{citations}

@ARTICLE{abazajian_2022,
       author = {{Abazajian}, Kevork and {Addison}, Graeme E. and {Adshead}, Peter and {Ahmed}, Zeeshan and {Akerib}, Daniel and {Ali}, Aamir and {Allen}, Steven W. and {Alonso}, David and {Alvarez}, Marcelo and {Amin}, Mustafa A. and {Anderson}, Adam and {Arnold}, Kam S. and {Ashton}, Peter and {Baccigalupi}, Carlo and {Bard}, Debbie and {Barkats}, Denis and {Barron}, Darcy and {Barry}, Peter S. and {Bartlett}, James G. and {Basu Thakur}, Ritoban and {Battaglia}, Nicholas and {Bean}, Rachel and {Bebek}, Chris and {Bender}, Amy N. and {Benson}, Bradford A. and {Bianchini}, Federico and {Bischoff}, Colin A. and {Bleem}, Lindsey and {Bock}, James J. and {Bocquet}, Sebastian and {Boddy}, Kimberly K. and {Richard Bond}, J. and {Borrill}, Julian and {Bouchet}, Fran{\c{c}}ois R. and {Brinckmann}, Thejs and {Brown}, Michael L. and {Bryan}, Sean and {Buza}, Victor and {Byrum}, Karen and {Hervias Caimapo}, Carlos and {Calabrese}, Erminia and {Calafut}, Victoria and {Caldwell}, Robert and {Carlstrom}, John E. and {Carron}, Julien and {Cecil}, Thomas and {Challinor}, Anthony and {Chang}, Clarence L. and {Chinone}, Yuji and {Sherry Cho}, Hsiao-Mei and {Cooray}, Asantha and {Coulton}, Will and {Crawford}, Thomas M. and {Crites}, Abigail and {Cukierman}, Ari and {Cyr-Racine}, Francis-Yan and {de Haan}, Tijmen and {Delabrouille}, Jacques and {Devlin}, Mark and {Di Valentino}, Eleonora and {Dierickx}, Marion and {Dobbs}, Matt and {Duff}, Shannon and {Dvorkin}, Cora and {Eimer}, Joseph and {Elleflot}, Tucker and {Errard}, Josquin and {Essinger-Hileman}, Thomas and {Fabbian}, Giulio and {Feng}, Chang and {Ferraro}, Simone and {Filippini}, Jeffrey P. and {Flauger}, Raphael and {Flaugher}, Brenna and {Fraisse}, Aurelien A. and {Frolov}, Andrei and {Galitzki}, Nicholas and {Gallardo}, Patricio A. and {Galli}, Silvia and {Ganga}, Ken and {Gerbino}, Martina and {Gluscevic}, Vera and {Goeckner-Wald}, Neil and {Green}, Daniel and {Grin}, Daniel and {Grohs}, Evan and {Gualtieri}, Riccardo and {Gudmundsson}, Jon E. and {Gullett}, Ian and {Gupta}, Nikhel and {Habib}, Salman and {Halpern}, Mark and {Halverson}, Nils W. and {Hanany}, Shaul and {Harrington}, Kathleen and {Hasegawa}, Masaya and {Hasselfield}, Matthew and {Hazumi}, Masashi and {Heitmann}, Katrin and {Henderson}, Shawn and {Hensley}, Brandon and {Hill}, Charles and {Colin Hill}, J. and {Hlo{\v{z}}ek}, Ren{\'e}e and {Patty Ho}, Shuay-Pwu and {Hoang}, Thuong and {Holder}, Gil and {Holzapfel}, William and {Hood}, John and {Hubmayr}, Johannes and {Huffenberger}, Kevin M. and {Hui}, Howard and {Irwin}, Kent and {Jeong}, Oliver and {Johnson}, Bradley R. and {Jones}, William C. and {Hwan Kang}, Jae and {Karkare}, Kirit S. and {Katayama}, Nobuhiko and {Keskitalo}, Reijo and {Kisner}, Theodore and {Knox}, Lloyd and {Koopman}, Brian J. and {Kosowsky}, Arthur and {Kovac}, John and {Kovetz}, Ely D. and {Kuhlmann}, Steve and {Kuo}, Chao-lin and {Kusaka}, Akito and {L{\"a}hteenm{\"a}ki}, Anne and {Lawrence}, Charles R. and {Lee}, Adrian T. and {Lewis}, Antony and {Li}, Dale and {Linder}, Eric and {Loverde}, Marilena and {Lowitz}, Amy and {Lubin}, Phil and {Madhavacheril}, Mathew S. and {Mantz}, Adam and {Marques}, Gabriela and {Matsuda}, Frederick and {Mauskopf}, Philip and {McCarrick}, Heather and {McMahon}, Jeffrey and {Daniel Meerburg}, P. and {Melin}, Jean-Baptiste and {Menanteau}, Felipe and {Meyers}, Joel and {Millea}, Marius and {Mohr}, Joseph and {Moncelsi}, Lorenzo and {Monzani}, Maria and {Mroczkowski}, Tony and {Mukherjee}, Suvodip and {Nagy}, Johanna and {Namikawa}, Toshiya and {Nati}, Federico and {Natoli}, Tyler and {Newburgh}, Laura and {Niemack}, Michael D. and {Nishino}, Haruki and {Nord}, Brian and {Novosad}, Valentine and {O'Brient}, Roger and {Padin}, Stephen and {Palladino}, Steven and {Partridge}, Bruce and {Petravick}, Don and {Pierpaoli}, Elena and {Pogosian}, Levon and {Prabhu}, Karthik and {Pryke}, Clement and {Puglisi}, Giuseppe and {Racine}, Benjamin and {Rahlin}, Alexandra and {Sathyanarayana Rao}, Mayuri and {Raveri}, Marco and {Reichardt}, Christian L. and {Remazeilles}, Mathieu and {Rocha}, Graca and {Roe}, Natalie A. and {Roy}, Anirban and {Ruhl}, John E. and {Salatino}, Maria and {Saliwanchik}, Benjamin and {Schaan}, Emmanuel and {Schillaci}, Alessandro and {Schmitt}, Benjamin and {Schmittfull}, Marcel M. and {Scott}, Douglas and {Sehgal}, Neelima and {Shandera}, Sarah and {Sherwin}, Blake D. and {Shirokoff}, Erik and {Simon}, Sara M. and {Slosar}, An{\v{z}}e and {Spergel}, David and {St. Germaine}, Tyler and {Staggs}, Suzanne T.},
        title = "{CMB-S4: Forecasting Constraints on Primordial Gravitational Waves}",
      journal = {\apj},
         year = 2022,
        month = feb,
       volume = {926},
       number = {1},
          eid = {54},
        pages = {54},
          doi = {10.3847/1538-4357/ac1596}
}

@ARTICLE{abitbol_2021,
       author = {{Abitbol}, Maximilian H. and {Alonso}, David and {Simon}, Sara M. and {Lashner}, Jack and {Crowley}, Kevin T. and {Ali}, Aamir M. and {Azzoni}, Susanna and {Baccigalupi}, Carlo and {Barron}, Darcy and {Brown}, Michael L. and {Calabrese}, Erminia and {Carron}, Julien and {Chinone}, Yuji and {Chluba}, Jens and {Coppi}, Gabriele and {Crowley}, Kevin D. and {Devlin}, Mark and {Dunkley}, Jo and {Errard}, Josquin and {Fanfani}, Valentina and {Galitzki}, Nicholas and {Gerbino}, Martina and {Hill}, J. Colin and {Johnson}, Bradley R. and {Jost}, Baptiste and {Keating}, Brian and {Krachmalnicoff}, Nicoletta and {Kusaka}, Akito and {Lee}, Adrian T. and {Louis}, Thibaut and {Madhavacheril}, Mathew S. and {McCarrick}, Heather and {McMahon}, Jeffrey and {Meerburg}, P. Daniel and {Nati}, Federico and {Nishino}, Haruki and {Page}, Lyman A. and {Poletti}, Davide and {Puglisi}, Giuseppe and {Randall}, Michael J. and {Rotti}, Aditya and {Spisak}, Jacob and {Suzuki}, Aritoki and {Teply}, Grant P. and {Verg{\`e}s}, Clara and {Wollack}, Edward J. and {Xu}, Zhilei and {Zannoni}, Mario},
        title = "{The Simons Observatory: gain, bandpass and polarization-angle calibration requirements for B-mode searches}",
      journal = {\jcap},
         year = 2021,
        month = may,
       volume = {2021},
       number = {5},
          eid = {032},
        pages = {032},
          doi = {10.1088/1475-7516/2021/05/032}
}

@ARTICLE{vielva_2022,
       author = {{Vielva}, P. and {Mart{\'\i}nez-Gonz{\'a}lez}, E. and {Casas}, F.~J. and {Matsumura}, T. and {Henrot-Versill{\'e}}, S. and {Komatsu}, E. and {Aumont}, J. and {Aurlien}, R. and {Baccigalupi}, C. and {Banday}, A.~J. and {Barreiro}, R.~B. and {Bartolo}, N. and {Calabrese}, E. and {Cheung}, K. and {Columbro}, F. and {Coppolecchia}, A. and {de Bernardis}, P. and {de Haan}, T. and {de la Hoz}, E. and {De Petris}, M. and {Della Torre}, S. and {Diego-Palazuelos}, P. and {Eriksen}, H.~K. and {Errard}, J. and {Finelli}, F. and {Franceschet}, C. and {Fuskeland}, U. and {Galloway}, M. and {Ganga}, K. and {Gervasi}, M. and {G{\'e}nova-Santos}, R.~T. and {Ghigna}, T. and {Gjerl{\o}w}, E. and {Gruppuso}, A. and {Hazumi}, M. and {Herranz}, D. and {Hivon}, E. and {Kohri}, K. and {Lamagna}, L. and {Leloup}, C. and {Macias-Perez}, J. and {Masi}, S. and {Matsuda}, F.~T. and {Morgante}, G. and {Nakano}, R. and {Nati}, F. and {Natoli}, P. and {Nerval}, S. and {Odagiri}, K. and {Oguri}, S. and {Pagano}, L. and {Paiella}, A. and {Paoletti}, D. and {Piacentini}, F. and {Polenta}, G. and {Puglisi}, G. and {Remazeilles}, M. and {Ritacco}, A. and {Rubino-Martin}, J.~A. and {Scott}, D. and {Sekimoto}, Y. and {Shiraishi}, M. and {Signorelli}, G. and {Takakura}, H. and {Tartari}, A. and {Thompson}, K.~L. and {Tristram}, M. and {Vacher}, L. and {Vittorio}, N. and {Wehus}, I.~K. and {Zannoni}, M. and {The LiteBIRD collaboration}},
        title = "{Polarization angle requirements for CMB B-mode experiments. Application to the LiteBIRD satellite}",
      journal = {\jcap},
         year = 2022,
        month = apr,
       volume = {2022},
       number = {4},
          eid = {029},
        pages = {029},
          doi = {10.1088/1475-7516/2022/04/029}
}

@article{Zaldarriaga1997,
  author  = {Zaldarriaga, Matias and Seljak, Uro{\v{s}}},
  title   = {All-sky analysis of polarization in the microwave background},
  journal = {\prd},
  year    = {1997},
  volume  = {55},
  number  = {4},
  pages   = {1830--1840},
  doi     = {10.1103/PhysRevD.55.1830},
  archivePrefix = {arXiv},
  eprint  = {astro-ph/9609170},
}

@article{Kamionkowski1997,
  author  = {Kamionkowski, Marc and Kosowsky, Arthur and Stebbins, Albert},
  title   = {Statistics of cosmic microwave background polarization},
  journal = {\prd},
  year    = {1997},
  volume  = {55},
  number  = {12},
  pages   = {7368--7388},
  doi     = {10.1103/PhysRevD.55.7368},
  archivePrefix = {arXiv},
  eprint  = {astro-ph/9611125},
}

@article{Ade2019,
  author  = {Ade, Peter and Aguirre, James and Ahmed, Zeeshan and Aiola, Simone and Ali, Aamir and {The Simons Observatory Collaboration}},
  title   = {The {Simons Observatory}: science goals and forecasts},
  journal = {\jcap},
  year    = {2019},
  volume  = {2019},
  number  = {02},
  pages   = {056},
  doi     = {10.1088/1475-7516/2019/02/056},
  archivePrefix = {arXiv},
  eprint  = {1808.07445},
}

@article{Kamionkowski2016,
  author  = {Kamionkowski, Marc and Kovetz, Ely D.},
  title   = {The Quest for {B} Modes from Inflationary Gravitational Waves},
  journal = {\araa},
  year    = {2016},
  volume  = {54},
  pages   = {227--269},
  doi     = {10.1146/annurev-astro-081915-023433},
}

@inproceedings{Hazumi2020,
  author    = {Hazumi, M. and Ade, P. A. R. and Adler, A. and Allys, E. and {LiteBIRD Collaboration}},
  title     = {{LiteBIRD} satellite: {JAXA's} new strategic {L-class} mission for all-sky surveys of cosmic microwave background polarization},
  booktitle = {Space Telescopes and Instrumentation 2020: Optical, Infrared, and Millimeter Wave},
  series    = {Proc. SPIE},
  volume    = {11443},
  pages     = {114432F},
  year      = {2020},
  doi       = {10.1117/12.2563050},
}

@article{ODea2007,
  author  = {O'Dea, Daniel and Challinor, Anthony and Johnson, Bradley R.},
  title   = {Systematic errors in cosmic microwave background polarization measurements},
  journal = {\mnras},
  year    = {2007},
  volume  = {376},
  number  = {4},
  pages   = {1767--1783},
  doi     = {10.1111/j.1365-2966.2007.11558.x},
  archivePrefix = {arXiv},
  eprint  = {astro-ph/0610361},
}

@article{Carroll1990,
  author  = {Carroll, Sean M. and Field, George B. and Jackiw, Roman},
  title   = {Limits on a Lorentz- and parity-violating modification of electrodynamics},
  journal = {\prd},
  year    = {1990},
  volume  = {41},
  number  = {4},
  pages   = {1231--1240},
  doi     = {10.1103/PhysRevD.41.1231},
}

@article{Keating2013,
  author  = {Keating, Brian G. and Shimon, Meir and Yadav, Amit P. S.},
  title   = {Self-Calibration of Cosmic Microwave Background Polarization Experiments},
  journal = {\apjl},
  year    = {2013},
  volume  = {762},
  number  = {2},
  pages   = {L23},
  doi     = {10.1088/2041-8205/762/2/L23},
  archivePrefix = {arXiv},
  eprint  = {1211.5734},
}

@article{Minami2019,
  author  = {Minami, Yuto and Ochi, Hiroki and Ichiki, Kiyotomo and Katayama, Nobuhiko and Komatsu, Eiichiro and Matsumura, Tomotake},
  title   = {Simultaneous determination of the cosmic birefringence and miscalibrated polarization angles from {CMB} experiments},
  journal = {PTEP},
  year    = {2019},
  volume  = {2019},
  number  = {8},
  pages   = {083E02},
  doi     = {10.1093/ptep/ptz079},
  archivePrefix = {arXiv},
  eprint  = {1904.12440},
}

@misc{Bock2006,
  author = {Bock, James and {DOE/NASA/NSF Interagency Task Force on CMB Research}},
  title  = {Task Force on Cosmic Microwave Background Research},
  year   = {2006},
  archivePrefix = {arXiv},
  eprint = {astro-ph/0604101},
}

@article{Johnson2015,
  author  = {Johnson, Bradley R. and Vourch, Camille J. and Drysdale, Thomas D. and Kalman, Andrew and Fujikawa, Sara and Keating, Brian and others},
  title   = {A {CubeSat} for Calibrating Ground-Based and Sub-Orbital Millimeter-Wave Polarimeters ({CalSat})},
  journal = {JAI},
  year    = {2015},
  volume  = {4},
  number  = {1},
  pages   = {1550007},
  doi     = {10.1142/S2251171715500075},
  archivePrefix = {arXiv},
  eprint  = {1505.07033},
}

@article{Krachmalnicoff2022,
  author  = {Krachmalnicoff, Nicoletta and Matsumura, Tomotake and de la Hoz, Ester and Basak, Susanta and Gruppuso, Alessandro and Minami, Yuto and others},
  title   = {In-flight polarization angle calibration for {LiteBIRD}: blind challenge and cosmological implications},
  journal = {\jcap},
  year    = {2022},
  volume  = {2022},
  number  = {01},
  pages   = {039},
  doi     = {10.1088/1475-7516/2022/01/039},
  archivePrefix = {arXiv},
  eprint  = {2111.09140},
}

@article{PlanckCollaboration2015,
  author  = {{Planck Collaboration} and Adam, R. and Ade, P. A. R. and Aghanim, N. and Arnaud, M. and others},
  title   = {{Planck} 2015 results. {VIII}. High Frequency Instrument data processing: Calibration and maps},
  journal = {\aap},
  year    = {2016},
  volume  = {594},
  pages   = {A8},
  doi     = {10.1051/0004-6361/201525820},
  archivePrefix = {arXiv},
  eprint  = {1502.01587},
}

@article{PlanckCollaboration2017,
  author  = {{Planck Collaboration} and Akrami, Y. and Ashdown, M. and Aumont, J. and Baccigalupi, C. and others},
  title   = {{Planck} intermediate results. {LII}. Planet flux densities},
  journal = {\aap},
  year    = {2017},
  volume  = {607},
  pages   = {A122},
  doi     = {10.1051/0004-6361/201630311},
}

@article{Aumont2020,
  author  = {Aumont, J. and Mac{\'\i}as-P{\'e}rez, J. F. and Ritacco, A. and Ponthieu, N. and Mangilli, A.},
  title   = {Absolute calibration of the polarisation angle for future {CMB} {B}-mode experiments from current and future measurements of the {Crab} nebula},
  journal = {\aap},
  year    = {2020},
  volume  = {634},
  pages   = {A100},
  doi     = {10.1051/0004-6361/201833504},
  archivePrefix = {arXiv},
  eprint  = {1911.03164},
}

@article{Weiland2011,
  author  = {Weiland, J. L. and Odegard, N. and Hill, R. S. and Wollack, E. and Hinshaw, G. and others},
  title   = {Seven-Year {Wilkinson Microwave Anisotropy Probe (WMAP)} Observations: Planets and Celestial Calibration Sources},
  journal = {\apjs},
  year    = {2011},
  volume  = {192},
  number  = {2},
  pages   = {19},
  doi     = {10.1088/0067-0049/192/2/19},
  archivePrefix = {arXiv},
  eprint  = {1001.4731},
}

@article{Lungu2022,
  author  = {Lungu, Marius and Storer, Emilie R. and Hasselfield, Matthew and Duivenvoorden, Adriaan J. and Calabrese, Erminia and others},
  title   = {The {Atacama Cosmology Telescope}: measurement and analysis of {1D} beams for {DR4}},
  journal = {\jcap},
  year    = {2022},
  volume  = {2022},
  number  = {05},
  pages   = {044},
  doi     = {10.1088/1475-7516/2022/05/044},
  archivePrefix = {arXiv},
  eprint  = {2112.12226},
}

@inproceedings{Dunner2024,
  author    = {D{\"u}nner, Rolando and Astori, Federico and Coppi, Gabriele and Eimer, Joseph and Errard, Josquin and Flux{\'a}, Juan and Li, Yunyang and Mezzanzanica, Noemi and Nati, Federico and Petroff, Matthew and Puddu, Roberto and Rojas, Mat{\'\i}as and Zannoni, Mario},
  title     = {{HoverCal} + {PoloCalC}: precise on-flight metrology performance results over {CLASS} telescope},
  booktitle = {Millimeter, Submillimeter, and Far-Infrared Detectors and Instrumentation for Astronomy XII},
  series    = {Proc. SPIE},
  volume    = {13102},
  pages     = {1310217},
  year      = {2024},
  doi       = {10.1117/12.3018469},
}

@article{Coppi2025,
  author  = {Coppi, Gabriele and Dachlythra, Nadia and Nati, Federico and D{\"u}nner-Planella, Rolando and Adler, Alexandre E. and Errard, Josquin and Galitzki, Nicholas and Li, Yunyang and Petroff, Matthew A. and Simon, Sara M. and Sang, Ema Tsang King and Aguilar, Amalia Villarrubia and Wollack, Edward J. and Zannoni, Mario},
  title   = {{PROTOCALC}, A {W}-band Polarized Calibrator for Cosmic Microwave Background Telescopes: Application to {Simons Observatory} and {CLASS}},
  journal = {\apjs},
  year    = {2025},
  volume  = {279},
  number  = {1},
  pages   = {30},
  doi     = {10.3847/1538-4365/adde5f},
  archivePrefix = {arXiv},
  eprint  = {2502.14473},
}

@article{Lang2010,
  author  = {Lang, Dustin and Hogg, David W. and Mierle, Keir and Blanton, Michael and Roweis, Sam},
  title   = {Astrometry.net: Blind Astrometric Calibration of Arbitrary Astronomical Images},
  journal = {\aj},
  year    = {2010},
  volume  = {139},
  number  = {5},
  pages   = {1782--1800},
  doi     = {10.1088/0004-6256/139/5/1782},
  archivePrefix = {arXiv},
  eprint  = {0910.2233},
}

@inproceedings{Shupe2005,
  author    = {Shupe, David L. and Moshir, Mehrdad and Li, Jing and Makovoz, David and Narron, Robert and Hook, Richard N.},
  title     = {The {SIP} Convention for Representing Distortion in {FITS} Image Headers},
  booktitle = {Astronomical Data Analysis Software and Systems XIV},
  series    = {ASP Conference Series},
  volume    = {347},
  pages     = {491--495},
  year      = {2005},
  editor    = {Shopbell, P. L. and Britton, M. C. and Ebert, R.},
}

@article{Bird1960,
  author  = {Bird, George R. and Parrish, Maxfield},
  title   = {The Wire Grid as a Near-Infrared Polarizer},
  journal = {JOSA},
  year    = {1960},
  volume  = {50},
  number  = {9},
  pages   = {886--891},
  doi     = {10.1364/JOSA.50.000886},
}

@article{Chen2025,
  author  = {Chen, Bing and Fu, Xiuhua and Liu, Xianzhu and Jiang, Huilin and others},
  title   = {A Review of Sub-Wavelength Wire Grid Polarizers and Their Development Trends},
  journal = {Photonics},
  year    = {2025},
  volume  = {12},
  number  = {11},
  pages   = {1046},
  doi     = {10.3390/photonics12111046},
}

@book{Hecht2017,
  author    = {Hecht, Eugene},
  title     = {Optics},
  edition   = {5th},
  publisher = {Pearson},
  address   = {Harlow, England},
  year      = {2017},
  isbn      = {978-1-292-09693-3},
}

@article{Casas2026,
  author  = {Casas, Francisco Javier and Mart{\'\i}nez-Gonz{\'a}lez, Enrique and Cubas, Javier and Pascual-Cisneros, Guillermo and S{\'a}nchez-Ramos, Irene and Castell{\'o}, Laura and Vielva, Patricio and Barreiro, Bel{\'e}n},
  title   = {{LEO-CalSat} for the calibration of {W}-band ground-based {CMB} polarization experiments},
  journal = {FrASS},
  year    = {2026},
  volume  = {13},
  pages   = {1671698},
  doi     = {10.3389/fspas.2026.1671698},
}

@book{Goodman2017,
  author    = {Goodman, Joseph W.},
  title     = {Introduction to Fourier Optics},
  edition   = {4th},
  publisher = {Macmillan Learning},
  address   = {New York},
  year      = {2017},
  isbn      = {9781319119164},
}

@article{2001PASP..113..622H,
  author  = {{Houde}, Martin and {Akeson}, Rachel L. and {Carlstrom}, John E. and {Lamb}, James W. and {Schleuning}, David A. and {Woody}, David P.},
  title   = {{Polarizing Grids, Their Assemblies, and Beams of Radiation}},
  journal = {\pasp},
  year    = {2001},
  month   = may,
  volume  = {113},
  number  = {783},
  pages   = {622--638},
  doi     = {10.1086/320287},
  archivePrefix = {arXiv},
  eprint  = {astro-ph/0311414},
}

@misc{bass2026tele_img_sim,
  author    = {Bass, Dillon},
  title     = {tele-img-sim: A simulation to model astrophotographs under diffraction and mounting instability},
  month     = oct,
  year      = {2026},
  version   = {v1.1},
  publisher = {Zenodo},
  doi       = {10.5281/zenodo.23420007},
  url       = {https://doi.org/10.5281/zenodo.23420007},
}

@misc{bass2026spectrangle,
  author    = {Bass, Dillon},
  title     = {{SpectrAngle}},
  month     = nov,
  year      = {2026},
  version   = {v1.2},
  publisher = {Zenodo},
  doi       = {10.5281/zenodo.24020974},
  url       = {https://doi.org/10.5281/zenodo.24020974},
}

@inproceedings{Voellmer2008,
  author    = {Voellmer, George M. and Bennett, Charles L. and Chuss, David T. and Eimer, Joseph R. and Hui, Hoi and Moseley, S. Harvey and Novak, Giles and Wollack, Edward J. and Zeng, Lingzhen},
  title     = {A large free-standing wire grid for microwave variable-delay polarization modulation},
  booktitle = {Ground-based and Airborne Instrumentation for Astronomy II},
  series    = {Proc. SPIE},
  volume    = {7014},
  pages     = {70142A},
  year      = {2008},
  doi       = {10.1117/12.787979},
}

@article{Matsuda2019,
  author  = {Matsuda, Frederick and Lowry, Lindsay and Suzuki, Aritoki and Aguilar F{\'a}undez, Mario and Arnold, Kam and Barron, Darcy and Bianchini, Federico and Cheung, Kolen and Chinone, Yuji and Elleflot, Tucker and Fabbian, Giulio and Goeckner-Wald, Neil and Hasegawa, Masaya and Kaneko, Daisuke and Katayama, Nobuhiko and Keating, Brian and Lee, Adrian T. and Navaroli, Martin and Nishino, Haruki and Paar, Hans and Puglisi, Giuseppe and Richards, Paul L. and Seibert, Joseph and Siritanasak, Praween and Tajima, Osamu and Takatori, Sayuri and Tsai, Calvin and Westbrook, Ben},
  title   = {The {POLARBEAR} Fourier Transform Spectrometer Calibrator and Spectroscopic Characterization of the {POLARBEAR} Instrument},
  journal = {RScI},
  volume  = {90},
  number  = {11},
  pages   = {115115},
  year    = {2019},
  doi     = {10.1063/1.5095160},
}

@article{GaiaDR3,
  author  = {{Gaia Collaboration} and Vallenari, A. and Brown, A. G. A. and Prusti, T. and de Bruijne, J. H. J. and Arenou, F. and Babusiaux, C. and Biermann, M. and others},
  title   = {{Gaia Data Release 3: Summary of the content and survey properties}},
  journal = {\aap},
  volume  = {674},
  pages   = {A1},
  year    = {2023},
  doi     = {10.1051/0004-6361/202243940},
  archivePrefix = {arXiv},
  eprint  = {2208.00211},
}

@article{Nati_2017,
  author  = {Nati, Federico and Devlin, Mark J. and Gerbino, Martina and Johnson, Bradley R. and Keating, Brian and Pagano, Luca and Teply, Grant},
  title   = {{POLOCALC}: A Novel Method to Measure the Absolute Polarization Orientation of the Cosmic Microwave Background},
  journal = {JAI},
  year    = {2017},
  volume  = {6},
  number  = {2},
  pages   = {1740008},
  doi     = {10.1142/S2251171717400086},
}

@article{Harvey1998,
  author  = {Harvey, James E. and Vernold, Cynthia L.},
  title   = {Description of diffraction grating behavior in direction cosine space},
  journal = {\ao},
  year    = {1998},
  volume  = {37},
  number  = {34},
  pages   = {8158--8160},
  doi     = {10.1364/AO.37.008158},
}

@techreport{Kaplan2005,
  author      = {Kaplan, G. H.},
  title       = {The {IAU} Resolutions on Astronomical Reference Systems, Time Scales, and Earth Rotation Models: Explanation and Implementation},
  institution = {U.S. Naval Observatory},
  type        = {USNO Circular},
  number      = {179},
  address     = {Washington, DC},
  year        = {2005},
  archivePrefix = {arXiv},
  eprint      = {astro-ph/0602086},
}

@article{Ritacco2024,
       author = {{Ritacco}, A. and {Bizzarri}, L. and {Savorgnano}, S. and
                 {Boulanger}, F. and {P{\'e}rault}, M. and {Treuttel}, J. and
                 {Morfin}, P. and {Catalano}, A. and {Darson}, D. and
                 {Ponthieu}, N. and {Feret}, A. and {Maffei}, B. and
                 {Chahadih}, A. and {Pisano}, G. and {Zannoni}, M. and
                 {Nati}, F. and {Mac{\'\i}as-P{\'e}rez}, J.~F. and
                 {Cuttaia}, F. and {Terenzi}, L. and {Monfardini}, A. and
                 {Calvo}, M. and {Murgia}, M. and {Ortu}, P. and
                 {Pisanu}, T. and {Aumont}, J. and {Errard}, J. and
                 {Leclercq}, S. and {Migliaccio}, M.},
        title = "{Absolute Reference for Microwave Polarization Experiments.
                 The COSMOCal Project and its Proof of Concept}",
      journal = {PASP},
         year = 2024,
        month = nov,
       volume = {136},
       number = {11},
          eid = {115001},
        pages = {115001},
          doi = {10.1088/1538-3873/ad8aed},
archivePrefix = {arXiv},
       eprint = {2405.12135},
 primaryClass = {astro-ph.IM}
}

@article{Savorgnano2026,
       author = {{Savorgnano}, S. and {Bizzarri}, L. and {Ritacco}, A. and {Aumont}, J. and {Boulanger}, F. and {Catalano}, A. and others},
        title = "{Proof of concept of the COSMOCal project at the IRAM 30m telescope}",
      journal = {arXiv e-prints},
         year = 2026,
        month = aug,
          eid = {arXiv:2608.05395},
        pages = {arXiv:2608.05395},
          doi = {10.48550/arXiv.2608.05395},
archivePrefix = {arXiv},
       eprint = {2608.05395},
 primaryClass = {astro-ph.IM}
}

@ARTICLE{Micheli2026,
       author = {{Micheli}, S. and {Ritacco}, A. and {Carones}, A. and others},
        title = "{Status and future development of the COSMOCal Project for absolute CMB polarization calibration}",
      journal = {arXiv e-prints},
         year = 2026,
        month = jul,
          eid = {arXiv:2607.12698},
        pages = {arXiv:2607.12698},
          doi = {10.48550/arXiv.2607.12698},
archivePrefix = {arXiv},
       eprint = {2607.12698},
 primaryClass = {astro-ph.IM}
}

@ARTICLE{Minami2020,
  author = {{Minami}, Y. and {Komatsu}, E.},
  title = "{New Extraction of the Cosmic Birefringence from the Planck 2018 Polarization Data}",
  journal = {PhRvL}, year = 2020, volume = {125}, eid = {221301}, pages = {221301},
  doi = {10.1103/PhysRevLett.125.221301}}

@ARTICLE{DiegoPalazuelos2022,
  author = {{Diego-Palazuelos}, P. and {Eskilt}, J.~R. and {Minami}, Y. and others},
  title = "{Cosmic Birefringence from the Planck Data Release 4}",
  journal = {PhRvL}, year = 2022, volume = {128}, eid = {091302}, pages = {091302},
  doi = {10.1103/PhysRevLett.128.091302}}

@ARTICLE{Murata2023,
  author = {{Murata}, M. and {Nakata}, H. and {Iijima}, K. and others},
  title = "{The Simons Observatory: A fully remote controlled calibration system with a sparse wire grid for cosmic microwave background telescopes}",
  journal = {RScI}, year = 2023, volume = {94}, eid = {124502}, pages = {124502},
  doi = {10.1063/5.0175099}}

@ARTICLE{Nakata2026,
  author = {{Nakata}, H. and {Adachi}, S. and {Yamada}, K. and others},
  title = "{The Simons Observatory: Detector Polarization Angle Calibration Using a Sparse Wire Grid with Initial Datasets of the Small-aperture Telescopes}",
  journal = {ApJS}, year = 2026, volume = {283}, eid = {78}, pages = {78},
  doi = {10.3847/1538-4365/ae4ddf}}

@INPROCEEDINGS{Cornelison2022,
  author = {{Cornelison}, J. and {Verg{\`e}s}, C. and {Ade}, P.~A.~R. and others},
  title = "{Improved polarization calibration of the BICEP3 CMB polarimeter at the South Pole}",
  booktitle = {Proc. SPIE}, year = 2022, volume = {12190}, pages = {121901X},
  doi = {10.1117/12.2620212}}

@ARTICLE{BK18angles2024,
       author = {{BICEP/Keck Collaboration} and {Ade}, P.~A.~R. and {Ahmed}, Z. and {Amiri}, M. and {Barkats}, D. and {Basu Thakur}, R. and {Bischoff}, C.~A. and {Beck}, D. and {Bock}, J.~J. and {Boenish}, H. and others},
        title = "{BICEP/Keck XVIII: Measurement of BICEP3 polarization angles and consequences for constraining cosmic birefringence and inflation}",
      journal = {PhRvD},
         year = 2025,
        month = mar,
       volume = {111},
       number = {6},
          eid = {063505},
        pages = {063505},
          doi = {10.1103/PhysRevD.111.063505},
archivePrefix = {arXiv},
       eprint = {2410.12089},
 primaryClass = {astro-ph.CO}
}

@article{numpy2020,
  author  = {Harris, Charles R. and Millman, K. Jarrod and van der Walt, St{\'e}fan J. and Gommers, Ralf and Virtanen, Pauli and Cournapeau, David and others},
  title   = {Array programming with {NumPy}},
  journal = {\nat},
  year    = {2020},
  volume  = {585},
  pages   = {357--362},
  doi     = {10.1038/s41586-020-2649-2},
}

@article{scipy2020,
  author  = {Virtanen, Pauli and Gommers, Ralf and Oliphant, Travis E. and Haberland, Matt and Reddy, Tyler and Cournapeau, David and others},
  title   = {{SciPy} 1.0: fundamental algorithms for scientific computing in {Python}},
  journal = {Nature Methods},
  year    = {2020},
  volume  = {17},
  pages   = {261--272},
  doi     = {10.1038/s41592-019-0686-2},
}

@article{matplotlib2007,
  author  = {Hunter, J. D.},
  title   = {Matplotlib: A {2D} Graphics Environment},
  journal = {Computing in Science \& Engineering},
  year    = {2007},
  volume  = {9},
  number  = {3},
  pages   = {90--95},
  doi     = {10.1109/MCSE.2007.55},
}

@misc{photutils2026,
  author    = {Bradley, Larry and Sip{\H{o}}cz, Brigitta M. and Robitaille, T. P. and Tollerud, E. J. and others},
  title     = {Photutils},
  month     = apr,
  year      = {2026},
  version   = {3.0.0},
  publisher = {Zenodo},
  doi       = {10.5281/zenodo.19636730},
  url       = {https://doi.org/10.5281/zenodo.19636730},
}

@ARTICLE{astropy:2022,
       author = {{Astropy Collaboration} and {Price-Whelan}, Adrian M. and {Lim}, Pey
       Lian and {Earl}, Nicholas and {Starkman}, Nathaniel and {Bradley}, Larry and
       {Shupe}, David L. and {Patil}, Aarya A. and {Corrales}, Lia and {Brasseur}, C.~E.
       and {N{\"o}the}, Maximilian and {Donath}, Axel and {Tollerud}, Erik and {Morris},
       Brett M. and {Ginsburg}, Adam and {Vaher}, Eero and {Weaver}, Benjamin A. and
       {Tocknell}, James and {Jamieson}, William and {van Kerkwijk}, Marten H. and
       {Robitaille}, Thomas P. and {Merry}, Bruce and {Bachetti}, Matteo and
       {G{\"u}nther}, H. Moritz and {Aldcroft}, Thomas L. and {Alvarado-Montes}, Jaime
       A. and {Archibald}, Anne M. and {B{\'o}di}, Attila and {Bapat}, Shreyas and
       {Barentsen}, Geert and {Baz{\'a}n}, Juanjo and {Biswas}, Manish and {Boquien},
       M{\'e}d{\'e}ric and {Burke}, D.~J. and {Cara}, Daria and {Cara}, Mihai and
       {Conroy}, Kyle E. and {Conseil}, Simon and {Craig}, Matthew W. and {Cross},
       Robert M. and {Cruz}, Kelle L. and {D'Eugenio}, Francesco and {Dencheva}, Nadia
       and {Devillepoix}, Hadrien A.~R. and {Dietrich}, J{\"o}rg P. and {Eigenbrot},
       Arthur Davis and {Erben}, Thomas and {Ferreira}, Leonardo and {Foreman-Mackey},
       Daniel and {Fox}, Ryan and {Freij}, Nabil and {Garg}, Suyog and {Geda}, Robel and
       {Glattly}, Lauren and {Gondhalekar}, Yash and {Gordon}, Karl D. and {Grant},
       David and {Greenfield}, Perry and {Groener}, Austen M. and {Guest}, Steve and
       {Gurovich}, Sebastian and {Handberg}, Rasmus and {Hart}, Akeem and
       {Hatfield-Dodds}, Zac and {Homeier}, Derek and {Hosseinzadeh}, Griffin and
       {Jenness}, Tim and {Jones}, Craig K. and {Joseph}, Prajwel and {Kalmbach}, J.
       Bryce and {Karamehmetoglu}, Emir and {Ka{\l}uszy{\'n}ski}, Miko{\l}aj and
       {Kelley}, Michael S.~P. and {Kern}, Nicholas and {Kerzendorf}, Wolfgang E. and
       {Koch}, Eric W. and {Kulumani}, Shankar and {Lee}, Antony and {Ly}, Chun and
       {Ma}, Zhiyuan and {MacBride}, Conor and {Maljaars}, Jakob M. and {Muna}, Demitri
       and {Murphy}, N.~A. and {Norman}, Henrik and {O'Steen}, Richard and {Oman}, Kyle
       A. and {Pacifici}, Camilla and {Pascual}, Sergio and {Pascual-Granado}, J. and
       {Patil}, Rohit R. and {Perren}, Gabriel I. and {Pickering}, Timothy E. and
       {Rastogi}, Tanuj and {Roulston}, Benjamin R. and {Ryan}, Daniel F. and {Rykoff},
       Eli S. and {Sabater}, Jose and {Sakurikar}, Parikshit and {Salgado}, Jes{\'u}s
       and {Sanghi}, Aniket and {Saunders}, Nicholas and {Savchenko}, Volodymyr and
       {Schwardt}, Ludwig and {Seifert-Eckert}, Michael and {Shih}, Albert Y. and
       {Jain}, Anany Shrey and {Shukla}, Gyanendra and {Sick}, Jonathan and {Simpson},
       Chris and {Singanamalla}, Sudheesh and {Singer}, Leo P. and {Singhal}, Jaladh and
       {Sinha}, Manodeep and {Sip{\H{o}}cz}, Brigitta M. and {Spitler}, Lee R. and
       {Stansby}, David and {Streicher}, Ole and {{\v{S}}umak}, Jani and {Swinbank},
       John D. and {Taranu}, Dan S. and {Tewary}, Nikita and {Tremblay}, Grant R. and
       {Val-Borro}, Miguel de and {Van Kooten}, Samuel J. and {Vasovi{\'c}}, Zlatan and
       {Verma}, Shresth and {de Miranda Cardoso}, Jos{\'e} Vin{\'\i}cius and {Williams},
       Peter K.~G. and {Wilson}, Tom J. and {Winkel}, Benjamin and {Wood-Vasey}, W.~M.
       and {Xue}, Rui and {Yoachim}, Peter and {Zhang}, Chen and {Zonca}, Andrea and
       {Astropy Project Contributors}}, title = "{The Astropy Project: Sustaining and
       Growing a Community-oriented Open-source Project and the Latest Major Release
       (v5.0) of the Core Package}",
      journal = {\apj},
         year = 2022,
        month = aug,
       volume = {935},
       number = {2},
          eid = {167},
        pages = {167},
          doi = {10.3847/1538-4357/ac7c74},
archivePrefix = {arXiv},
       eprint = {2206.14220},
 primaryClass = {astro-ph.IM},
       adsurl = {https://ui.adsabs.harvard.edu/abs/2022ApJ...935..167A}
}

@ARTICLE{astropy:2018,
   author = {{Astropy Collaboration} and {Price-Whelan}, A.~M. and {Sip{\H o}cz}, B.~M. and
	{G{\"u}nther}, H.~M. and {Lim}, P.~L. and {Crawford}, S.~M. and
	{Conseil}, S. and {Shupe}, D.~L. and {Craig}, M.~W. and {Dencheva}, N. and
	{Ginsburg}, A. and {VanderPlas}, J.~T. and {Bradley}, L.~D. and
	{P{\'e}rez-Su{\'a}rez}, D. and {de Val-Borro}, M. and {Paper Contributors}, (. and
	{Aldcroft}, T.~L. and {Cruz}, K.~L. and {Robitaille}, T.~P. and
	{Tollerud}, E.~J. and {Coordination Committee}, (. and {Ardelean}, C. and
	{Babej}, T. and {Bach}, Y.~P. and {Bachetti}, M. and {Bakanov}, A.~V. and
	{Bamford}, S.~P. and {Barentsen}, G. and {Barmby}, P. and {Baumbach}, A. and
	{Berry}, K.~L. and {Biscani}, F. and {Boquien}, M. and {Bostroem}, K.~A. and
	{Bouma}, L.~G. and {Brammer}, G.~B. and {Bray}, E.~M. and {Breytenbach}, H. and
	{Buddelmeijer}, H. and {Burke}, D.~J. and {Calderone}, G. and
	{Cano Rodr{\'{\i}}guez}, J.~L. and {Cara}, M. and {Cardoso}, J.~V.~M. and
	{Cheedella}, S. and {Copin}, Y. and {Corrales}, L. and {Crichton}, D. and
	{D{\'A}vella}, D. and {Deil}, C. and {Depagne}, {\'E}. and
	{Dietrich}, J.~P. and {Donath}, A. and {Droettboom}, M. and
	{Earl}, N. and {Erben}, T. and {Fabbro}, S. and {Ferreira}, L.~A. and
	{Finethy}, T. and {Fox}, R.~T. and {Garrison}, L.~H. and {Gibbons}, S.~L.~J. and
	{Goldstein}, D.~A. and {Gommers}, R. and {Greco}, J.~P. and
	{Greenfield}, P. and {Groener}, A.~M. and {Grollier}, F. and
	{Hagen}, A. and {Hirst}, P. and {Homeier}, D. and {Horton}, A.~J. and
	{Hosseinzadeh}, G. and {Hu}, L. and {Hunkeler}, J.~S. and {Ivezi{\'c}}, {\v Z}. and
	{Jain}, A. and {Jenness}, T. and {Kanarek}, G. and {Kendrew}, S. and
	{Kern}, N.~S. and {Kerzendorf}, W.~E. and {Khvalko}, A. and
	{King}, J. and {Kirkby}, D. and {Kulkarni}, A.~M. and {Kumar}, A. and
	{Lee}, A. and {Lenz}, D. and {Littlefair}, S.~P. and {Ma}, Z. and
	{Macleod}, D.~M. and {Mastropietro}, M. and {McCully}, C. and
	{Montagnac}, S. and {Morris}, B.~M. and {Mueller}, M. and {Mumford}, S.~J. and
	{Muna}, D. and {Murphy}, N.~A. and {Nelson}, S. and {Nguyen}, G.~H. and
	{Ninan}, J.~P. and {N{\"o}the}, M. and {Ogaz}, S. and {Oh}, S. and
	{Parejko}, J.~K. and {Parley}, N. and {Pascual}, S. and {Patil}, R. and
	{Patil}, A.~A. and {Plunkett}, A.~L. and {Prochaska}, J.~X. and
	{Rastogi}, T. and {Reddy Janga}, V. and {Sabater}, J. and {Sakurikar}, P. and
	{Seifert}, M. and {Sherbert}, L.~E. and {Sherwood-Taylor}, H. and
	{Shih}, A.~Y. and {Sick}, J. and {Silbiger}, M.~T. and {Singanamalla}, S. and
	{Singer}, L.~P. and {Sladen}, P.~H. and {Sooley}, K.~A. and
	{Sornarajah}, S. and {Streicher}, O. and {Teuben}, P. and {Thomas}, S.~W. and
	{Tremblay}, G.~R. and {Turner}, J.~E.~H. and {Terr{\'o}n}, V. and
	{van Kerkwijk}, M.~H. and {de la Vega}, A. and {Watkins}, L.~L. and
	{Weaver}, B.~A. and {Whitmore}, J.~B. and {Woillez}, J. and
	{Zabalza}, V. and {Contributors}, (.},
    title = "{The Astropy Project: Building an Open-science Project and Status of the v2.0 Core Package}",
  journal = {\aj},
archivePrefix = "arXiv",
   eprint = {1801.02634},
 primaryClass = "astro-ph.IM",
     year = 2018,
    month = sep,
   volume = 156,
      eid = {123},
    pages = {123},
      doi = {10.3847/1538-3881/aabc4f},
   adsurl = {https://ui.adsabs.harvard.edu/abs/2018AJ....156..123T}
}

@ARTICLE{astropy:2013,
   author = {{Astropy Collaboration} and {Robitaille}, T.~P. and {Tollerud}, E.~J. and
    {Greenfield}, P. and {Droettboom}, M. and {Bray}, E. and {Aldcroft}, T. and
    {Davis}, M. and {Ginsburg}, A. and {Price-Whelan}, A.~M. and
    {Kerzendorf}, W.~E. and {Conley}, A. and {Crighton}, N. and
    {Barbary}, K. and {Muna}, D. and {Ferguson}, H. and {Grollier}, F. and
    {Parikh}, M.~M. and {Nair}, P.~H. and {Unther}, H.~M. and {Deil}, C. and
    {Woillez}, J. and {Conseil}, S. and {Kramer}, R. and {Turner}, J.~E.~H. and
    {Singer}, L. and {Fox}, R. and {Weaver}, B.~A. and {Zabalza}, V. and
    {Edwards}, Z.~I. and {Azalee Bostroem}, K. and {Burke}, D.~J. and
    {Casey}, A.~R. and {Crawford}, S.~M. and {Dencheva}, N. and
    {Ely}, J. and {Jenness}, T. and {Labrie}, K. and {Lian Lim}, P. and
    {Pierfederici}, F. and {Pontzen}, A. and {Ptak}, A. and {Refsdal}, B. and
    {Servillat}, M. and {Streicher}, O.},
    title = "{Astropy: A community Python package for astronomy}",
  journal = {\aap},
     year = 2013,
    month = oct,
   volume = 558,
      eid = {A33},
    pages = {A33},
      doi = {10.1051/0004-6361/201322068},
   adsurl = {https://ui.adsabs.harvard.edu/abs/2013A%26A...558A..33A}
}

@ARTICLE{BK18,
  author  = {{BICEP/Keck Collaboration} and Ade, P.~A.~R. and others},
  title   = "{Improved Constraints on Primordial Gravitational Waves using Planck,
             WMAP, and BICEP/Keck Observations through the 2018 Observing Season}",
  journal = {Phys. Rev. Lett.}, year = 2021, volume = 127, pages = {151301},
  doi     = {10.1103/PhysRevLett.127.151301}}

@INPROCEEDINGS{Bryan2018,
  author    = {Bryan, S.~A. and Simon, S.~M. and Gerbino, M. and others},
  title     = "{Development of calibration strategies for the Simons Observatory}",
  booktitle = {Proc. SPIE}, volume = 10708, pages = {1070840}, year = 2018,
  doi       = {10.1117/12.2313832}}
\bibliographystyle{aasjournalv7.1}

\end{document}